\documentclass[11pt,a4paper]{article}
\pdfoutput=1

\usepackage{jheppub}
\usepackage{rotating}

\usepackage{wrapfig}
\usepackage{bbm}
\usepackage{amsfonts}
\usepackage{amsmath}
\usepackage{slashed}
\usepackage{amssymb}
\usepackage{latexsym}
\usepackage{multirow}
\usepackage{hyperref}
\usepackage{enumitem}
\hypersetup{colorlinks=true,citecolor=red,linkcolor=magenta,urlcolor=blue}
\usepackage[caption=false]{subfig}
\usepackage{relsize}
\usepackage{booktabs}
\usepackage{cleveref}
\usepackage{fancyhdr}
\usepackage{float}
 \usepackage{subfig}
\usepackage{graphicx}
\usepackage{microtype}
\usepackage{tabularx}
\usepackage{xcolor}
\usepackage{tensor}
\usepackage[bottom]{footmisc}

\def\reply#1{\textcolor{black}{#1}}

\newcommand{\ep}{\varepsilon}

\renewcommand{\arraystretch}{1.3}

\newcolumntype{C}{>{$}c<{$}}
\newcolumntype{L}{>{$}l<{$}}
\newcolumntype{R}{>{$}r<{$}}

\newcommand{\DeltaUV}{\Delta^{\text{UV}}}

\newcommand{\sigaz}{\Sigma^T_{AZ}}
\newcommand{\sigza}{\Sigma^T_{ZA}}

\newcommand{\MSb}{$\overline{\text{MS}}$}
\newcommand{\dmw}{\delta M_W^2}
\newcommand{\dmz}{\delta M_Z^2}
\newcommand{\dze}{\delta Z_e}
\newcommand{\dsw}{\delta s_W}
\newcommand{\dcw}{\delta c_W}
\newcommand{\dzw}{\delta Z_{W}}
\newcommand{\dzaa}{\delta Z_{AA}}
\newcommand{\dzaz}{\delta Z_{AZ}}
\newcommand{\dzza}{\delta Z_{ZA}}
\newcommand{\dzzz}{\delta Z_{ZZ}}
\newcommand{\da}{\delta a}
\newcommand{\dz}{\delta\zeta}

\newcommand{\mg}{\textsc{MadGraph5\_}aMC@NLO}

\title{Quartic Gauge-Higgs couplings: Constraints and Future Directions}
\author[a]{Anisha,}
\author[a]{Oliver Atkinson,}
\author[a]{Akanksha Bhardwaj,} 
\author[a]{Christoph Englert,} 
\author[a]{Panagiotis Stylianou} 
\affiliation[a]{School of Physics \& Astronomy, University of Glasgow, Glasgow G12 8QQ, United Kingdom}
\emailAdd{anisha@glasgow.ac.uk}
\emailAdd{o.atkinson.1@research.gla.ac.uk}
\emailAdd{akanksha.bhardwaj@glasgow.ac.uk}
\emailAdd{christoph.englert@glasgow.ac.uk}
\emailAdd{p.stylianou.1@research.gla.ac.uk}

\abstract{Constraints on quartic interactions of the Higgs boson with gauge bosons have been obtained by the experimental LHC collaborations focussing on the so-called $\kappa$ framework of flat rescalings of SM-like interactions in weak boson fusion (WBF) Higgs pair production. While such approaches are admissible to obtain a qualitative picture of consistency with the SM when the statistical yield is low, once more statistics become available a more theoretically consistent framework of limit setting is desirable. Reviewing the constraints provided at the Large Hadron Collider, we first show that these limits are robust when considered in a leading order context. Turning to radiative corrections, we demonstrate the limitations of this approach in the SM, and by adopting Higgs effective field theory techniques, we clarify the sensitivity from single Higgs measurements to rescalings of quartic Higgs-gauge couplings. We then discuss avenues for sensitivity improvements of WBF analyses employing Graph Neural Networks to combat the large contributing backgrounds.}
\begin{document}
%%%%%%%%%%%%%%%%%%%
\maketitle
%%%%%%%%%%%%%%%%%%%
%%%%%%%%%%%%%%%%%%%
\allowdisplaybreaks
\section{Introduction}
\label{sec:intro}
%%%%%%%%%%%%%%%%%%%
Ten years into the Higgs characterisation programme at the Large Hadron Collider (LHC), many couplings of the Higgs boson to other SM matter have been shown to largely follow the Standard Model (SM) expectation. Given the insights from electroweak fits before the Higgs discovery in 2012, this was perhaps anticipated for the gauge sector~\cite{Baak:2011ze,Falkowski:2013dza}. The observation of $H\to \gamma\gamma$ with a rate in agreement with the SM prediction indicated alignment of the fermion-Higgs interactions as part of the Higgs discovery for the top quark. Later analyses in other fermion decay modes of the Higgs boson~\cite{ATLAS:2021qou,ATLAS:2022yrq,CMS:2022kdi,CMS:2017odg} have further established the Higgs boson as SM-like.

Couplings that have received less attention, also because corresponding searches are statistics-limited, are the quartic interactions of the Higgs boson with gauge fields. Assuming a weak doublet-like character of the Higgs boson, these interactions are fully correlated with trilinear Higgs-gauge ($HWW$, $HZZ$) interactions due to gauge invariance in the SM. Searches for anomalous quartic interactions in the so-called $\kappa$ framework~\cite{LHCHiggsCrossSectionWorkingGroup:2012nn} are therefore theoretically cumbersome, albeit instructive when the statistical sensitivity is low (similar to $\kappa$-like analyses of single Higgs observables after the Higgs boson's discovery).

In the expansion of the Higgs boson around its vacuum expectation value (vev), the gauge boson masses are a property of the non-linear $SU(2)_L\times U(1)_Y$ electroweak symmetry realisation, whilst the Higgs bosons' couplings to $W$ and $Z$ bosons probe the alignment of quantum fluctuations around this vev. These interactions carry important information as any departure from the SM expectation immediately implies perturbative unitarity violation at a scale above the vev~\cite{Lee:1977eg,Chanowitz:1978mv,Abu-Ajamieh:2020yqi}, which acts as a strong indicator of a new scale of physics beyond the SM (BSM). 

Quartic interactions $\sim H^2 V^2$ ($V=W,Z$), while being less relevant from a perturbative unitarity perspective, still carry important information when $HVV$ couplings (dis)agree with the SM: Different BSM scenarios show similar alignment of single Higgs interactions with the vev. In the gauge sector this degeneracy is only lifted by considering quartic interactions. An example for this is custodial singlet mixing~\cite{Binoth:1996au,Schabinger:2005ei,Patt:2006fw,Englert:2011yb} compared to minimal compositeness Higgs models (MCHMs). In the former case Higgs interactions are modified by a characteristic mixing angle $\sim \cos\theta$ whereas in, e.g., MCHM5~\cite{Contino:2006qr}, we obtain a modification of $\sim \sqrt{1-v^2/f^2}$~\cite{Contino:2006qr,Grober:2010yv,Gillioz:2012se} (where $v\simeq 246~\text{GeV}$ is the vev and $f\gtrsim v$ is the decay constant of the Callen-Coleman-Wess-Zumino (CCWZ)~\cite{Coleman:1969sm,Callan:1969sn} construction). On the sole basis of a precision analysis of the $HWW,HZZ$ interactions, it is therefore impossible to discriminate between Higgs mixing and Higgs compositeness as we can always understand $\cos^2\theta = 1-v^2/f^2$ (note that custodial symmetry is sufficiently conserved in both models~\cite{Agashe:2006at}). Only when we consider the curvature of the Higgs interactions is this ambiguity lifted in the gauge sector: in the custodial singlet mixing scenario, the $H^2V^2$ vertex is modified by $\cos^2 \theta \neq 1-2v^2/f^2$, where the latter is the result in MCHM5.

Targeted analyses for flat rescalings
\begin{equation}
\kappa_{2V} = {g_{HHVV}\over g^{\text{SM}}_{HHVV}}\,,
\end{equation}
in weak boson fusion (WBF) Higgs pair production show that this channel will be statistics-limited at the high-luminosity (HL) frontier~\cite{Dolan:2013rja,Dolan:2015zja,Bishara:2016kjn,Kilian:2018bhs,Arganda:2018ftn}. To some degree, this is similar to investigations of the Higgs boson's self-coupling; statistics is also a limiting factor at the HL-LHC for rescalings of the trilinear coupling $\kappa_3$~\cite{Dolan:2012rv}. To partially address this issue, single Higgs probes have been suggested~\cite{McCullough:2013rea,Maltoni:2017ims,Degrassi:2016wml,DiVita:2017eyz,Voigt:2017vfz,Englert:2019eyl} to add sensitivity. The comparison of single-Higgs probes (or even electroweak precision data~\cite{Kribs:2017znd,Englert:2020gcp}) to rescalings of the Higgs self-coupling $\kappa_3$ should, however, also be understood as a theoretical figure of merit related to the trustworthiness of the obtained constraints as different orders in the perturbative (effective field theory) series expansion are tensioned against each other.

Given that WBF Higgs pair production is significantly smaller than inclusive Higgs pair production via gluon fusion~\cite{LHCHiggsCrossSectionWorkingGroup:2016ypw}, similar issues of $\kappa_{2V}$ do occur. It is the purpose of this work to clarify these questions in the light of existing analyses pursued by the ATLAS and CMS experiments~\cite{ATLAS:2020jgy,ATLAS:2022ycx,CMS:2022hgz}. We show that this is achieved by understanding the SM as a particular parameter point of the electroweak chiral (or Higgs Effective Theory, HEFT) Lagrangian. This enables us to consistently modify the quartic Higgs-gauge interactions away from the SM expectation with theoretically consistent implications for single Higgs observables.
In particular, we will focus on radiative corrections to the $H\to VV$ rates, which we tension against direct sensitivity obtained from WBF Higgs pair analyses. To formulate an as-good-as-possible direct search limit, we employ machine learning techniques to the signal-background discrimination to show that significant improvements are possible beyond the existing, more traditional techniques discussed in the literature.

This work is organised as follows: In Sec.~\ref{sec:kap2v}, we review the theoretical baseline of $\kappa_{2V}$ analyses in the SM. We consider electroweak precision constraints from oblique corrections and derive unitarity constraints from a partial wave analysis, clarifying the validity of the results of~\cite{ATLAS:2020jgy,ATLAS:2022ycx,CMS:2022hgz} viewed as a leading order signal analysis. We show that a $\kappa_{2V}$ analysis cannot be theoretically defended when we consider quantum corrections in the SM. To overcome this limitation, we turn to the electroweak chiral Lagrangian in Sec.~\ref{sec:heft}, which provides the theoretically consistent playground to discuss correlations between single and double Higgs modifiers at a given order of the amplitudes' loop expansion. Performing the next-to-leading order calculation of $H\to ZZ^\ast,WW^\ast$ alongside $H\to \gamma\gamma, \gamma Z$, we discuss the current and expected sensitivity of single Higgs data to $\kappa_{2V}$ coupling modifiers. This highlights the direct sensitivity provided in weak boson fusion $HHjj$ analysis as the sensitive tool to constrain $\kappa_{2V}$ in the future (for a recent discussion at $e^+e^-$ machines see~Ref.~\cite{Domenech:2022uud}). Any improvement beyond increasing statistics of current strategies~\cite{ATLAS:2020jgy,ATLAS:2022ycx,CMS:2022hgz} will therefore be directly correlated with an improved understanding of the Higgs boson's TeV scale characteristics. To this end, we show that the application of state-of-the-art Graph Neural Network (GNN) techniques~\cite{zhou2018graph,9046288} to $pp \to HHjj$ final states can greatly reduce the dominant backgrounds of such searches, thus leading to a significant sensitivity enhancement to $\kappa_{2V}$. We summarise and conclude in Sec.~\ref{sec:conc}.

%%%%%%%%%%%%%%%%%%%
\section{Vices and Virtues of $\kappa_{2V}$}
\label{sec:kap2v}
%%%%%%%%%%%%%%%%%%%
We consider first the effect of flatly rescaled quartic Higgs-gauge couplings to electroweak precision constraints. Here, we particularly focus on oblique corrections~\cite{Peskin:1990zt,Peskin:1991sw} and their sensitivity to coupling modifications of {\emph{only}} the physical Higgs boson (with mass $M_H\simeq 125~\text{GeV}$). These interactions are not associated with momentum dependencies at one-loop order so we directly obtain
\begin{equation}
\Delta S = \Delta U= 0 \,,
\end{equation}
(we do not consider single Higgs modifiers at this point). The $T$ parameter is sensitive to the custodial isospin violation for $\kappa_{2W}\neq \kappa_{2Z}$
\begin{equation}
\label{eq:tparam}
\Delta T = { \kappa_{2Z}^2 - \kappa_{2W}^2 \over 16 \pi } {M_H^2 \over M_W^2 s_W^2}  \log {\Lambda^2 \over M_H^2}\,,
\end{equation}
with $\Lambda$ as the UV cutoff regulator for demonstration purposes.\footnote{In the following, we denote the sine and cosine of the Weinberg angle with $s_W,c_W$. The $W$ and $Z$ masses are labelled with $M_W,M_Z$, respectively.} This means that for large cutoffs of order 10~TeV we need to ensure $\kappa_{2W}\simeq \kappa_{2Z}$ at the 1.5\% level in order to not violate present constraints~\cite{Baak:2014ora} whilst UV engineering interactions that dominantly source custodial isospin violation through the quartic gauge-Higgs interactions. The latter is naively possible in the $\kappa$ framework, Eq.~\eqref{eq:tparam}, but rather unlikely when considered in the context of more realistic SM extensions, some of which we have alluded to above.
On the other hand, by identifying $\kappa_{2W} = \kappa_{2Z} = \kappa_{2V}$, constraints from oblique corrections are removed at the considered one-loop level. This is the identification we will use in the following; electroweak precision results suggest that the Higgs boson should be considered as a custodial iso-singlet state, which is the hypothesis on which the electroweak chiral Lagrangian is built.

%%%%%%%%%%%%%%%%%%%%%%
\begin{figure}[!t]
\centering \includegraphics[width=9cm]{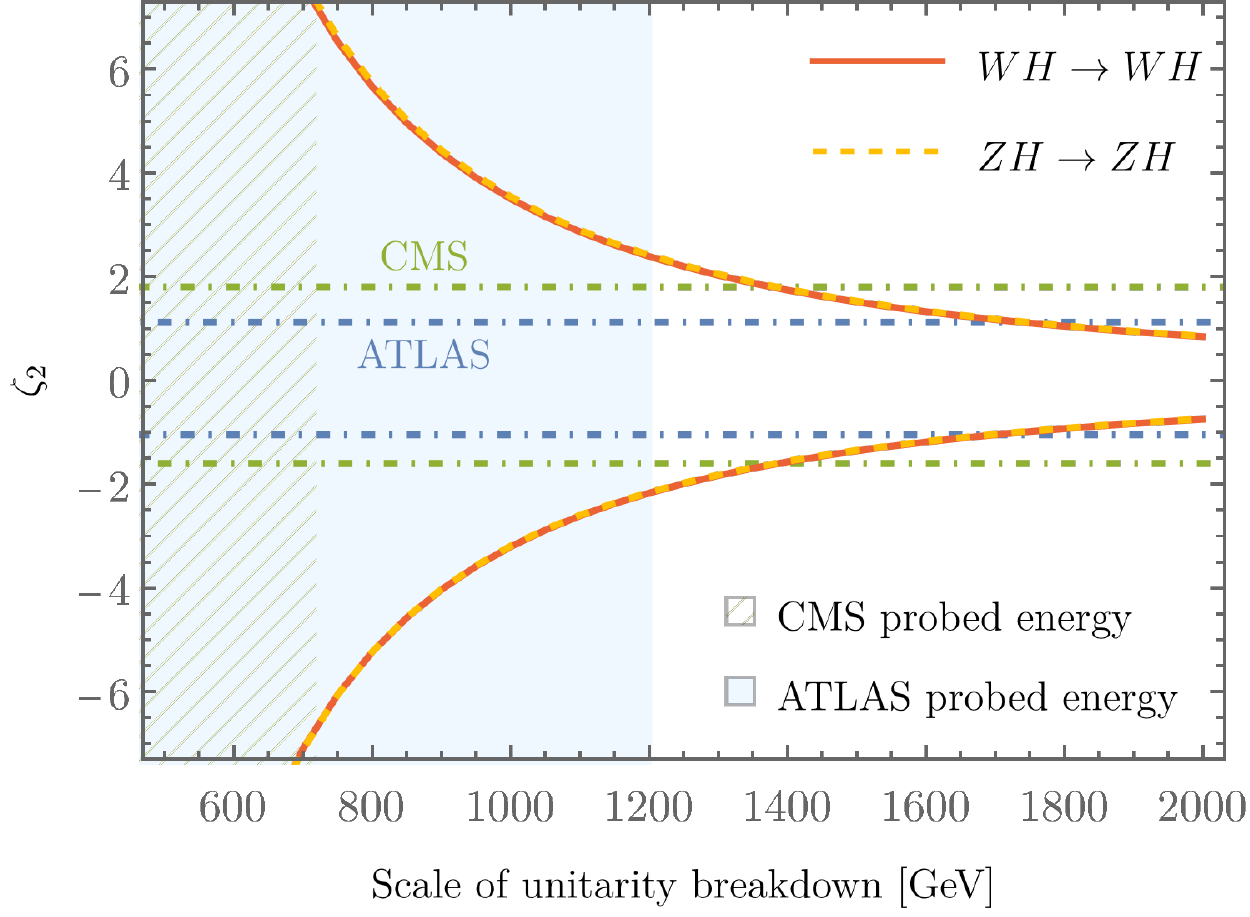}
	\caption{\label{fig:unitarity} Perturbative unitarity bounds on $\kappa_{2 V} = 1 + \zeta_{2}$ at different scales from the $V H \to V H$ channels. The expected $95$\% CL constraints from ATLAS~\cite{ATLAS:2022ycx} $\kappa_{2 V} \in \left[-0.05, 2.12\right]$, and from CMS~\cite{CMS:2022hgz} $\kappa_{2 V} \in \left[-0.6, 2.8\right]$ are also overlaid. The ATLAS results are obtained from the $HH \to b \bar{b} b \bar{b}$ final state probing energies in the shaded region $m_{HH} \leq 1200$~GeV, while CMS probed up to $m_{HH} \sim 720$~GeV with the $b \bar{b} \tau \bar{\tau}$ final state.}
\end{figure}
%%%%%%%%%%%%%%%%%%%%%%

Next, we turn to unitarity constraints. The partial waves of $i_1 i_2 \to f_1 f_2$ scattering process for angular momentum $J$ are given by~(see Ref.~\cite{Jacob:1959at})
\begin{equation}
a_{fi}^J = {\beta^{1/4}(s,m_{f_1}^2,m_{f_1}^2)\beta^{1/4}(s,m_{i_1}^2,m_{i_1}^2) \over 32\pi s}\int\limits_{-1}^1\text{d}\cos\theta \, {\cal{D}}^J_{\mu_i\mu_f} \,{\cal{M}}(s,\cos\theta)\,,
\end{equation}
(modulo factors of $1/\sqrt{2}$ for identical particles in the initial or final state) where we denote $\sqrt{s}$ as the centre-of-mass energy, the scattering angle in the centre-of-mass system is $\theta$, and
\begin{equation}
\beta(x, y, z) = x^2 + y^2 + z^2 - 2xy - 2yz - 2xz\,.
\end{equation}
The ${\cal{D}}^J_{\mu_i\mu_f}$ are the Wigner functions of the Jacob-Wick expansion~\cite{Jacob:1959at} related to the helicities of the initial and final states $\mu_i = {\lambda}_{i_1} - {\lambda}_{i_2}$, $\mu_f={\lambda}_{f_1}  - {\lambda}_{f_2}$. Unitarity of the $S$ matrix then translates for $f=i$ to the familiar conditions
\begin{equation}
\text{Re}|a_{ii}^j| \leq {1\over 2}\quad \hbox{and} \quad \text{Im} | a_{ii}^J | \leq 1\,.
\end{equation}
The process relevant for $\kappa_{2V}$ constraints is longitudinal $HV_L \to HV_L$ scattering, so that the Wigner functions reduce to the Legendre polynomials, with the $J=0$ partial wave providing the dominant constraint. This is shown in Fig.~\ref{fig:unitarity}, where we define
\begin{equation}
\kappa_{2V} = 1+ \zeta_{2}\,.
\end{equation}
We also include the LHC results from the ATLAS and CMS experiments of Refs.~\cite{ATLAS:2020jgy,ATLAS:2022ycx,CMS:2022hgz}. As can be seen, the unitarity constraints are relatively loose. The measurements, taken as a leading order result, can be considered perturbatively consistent as the energy scales explored by these analyses do not probe the unitarity cutoff of the leading order theory.

The discussion so far suggests that $\kappa_{2V}$ is indeed relatively unconstrained when naively considered as a free parameter in SM-like analyses. However, beyond tree-level, a choice of $\kappa_{2V}\neq 1$ leads to incurable shortfalls within the SM. For instance, when considering electroweak corrections to $H\to Z Z^\ast$ in general $R_\xi$ gauge~\cite{Fujikawa:1972fe}, we obtain a counter term amplitude (${\cal{M}}_{\text{CT}}$)-renormalised\footnote{We employ the on-shell renormalisation scheme here; further details are given in Sec.~\ref{sec:heft}.} virtual amplitude (neglecting fermion contributions and considering $M_H>2 M_Z$ for convenience)
\begin{multline}
\label{eq:oneloopc2v}
	{\cal{M}}^{\text{virt}} + {\cal{M}}_{\text{CT}}  = - {\alpha  \over 32\pi }{e \over M_W s_W^5 c_W^2} \left(M_H^2 + 2 M_Z^2 s_W^2 \{3+\xi_Z\} \right) \\ \times \Delta^{\text{UV}}(\mu^2,M_Z^2)\, \zeta_{2} \, [{\epsilon}^\mu(Z_1) {\epsilon}_\mu(Z_2) ]^\ast + {\cal{O}}(\ep^0)\,.
\end{multline}
$\epsilon(Z_i)$ are the transverse polarisations of the $Z$ bosons (related to the decay current $Z\to f\bar f$ for massless fermions) and 
\begin{equation}
\label{eq:deltadef}
\Delta^{\text{UV}}(\mu^2,M^2)= \left({4 \pi \mu^2 \over M^2}\right)^\ep {\Gamma(1+\ep)\over \ep} = {1\over \ep} - \gamma_E + \log(4\pi) + \log{\mu^2\over M^2} + {\cal{O}}(\ep) \,,
\end{equation}
is the usual $\overline {\text{MS}}$-related divergence in dimensional regularisation $D=4-2\ep$~\cite{tHooft:1972tcz}. 
This result is qualitatively not surprising: As we have modified without much care only a single part of the gauge-invariant $SU(2)_L\times U(1)_Y$ kinetic Higgs field term 
\begin{equation}
D_\mu\Phi^\dagger D^\mu \Phi \supset - g_{HZZ}^{\text{SM}} H Z^\mu Z_\mu -  g_{HHZZ}^{\text{SM}} H^2 Z^\mu Z_\mu \,,
\end{equation}
 ($\Phi=(\sqrt{2}G^+,v+H+iG^0)^T/\sqrt{2}$ being the Higgs doublet) rather than the whole term, we have lost gauge invariance entirely. This is highlighted by the non-vanishing $\xi_Z$ gauge fixing parameter dependence of the $H\to ZZ$ decay. $\zeta_{2}\neq 0$ has also induced singularities that are not mended by a renormalisation of the bare $HZZ$ SM interactions (included in Eq.~\eqref{eq:oneloopc2v}).\footnote{Considering off-shell $Z$ bosons for kinematics $M_H<2M_Z$ leads to similar issues alongside the appearance of additional divergences $\sim \xi_W$.} Again this is expected; radiative $\zeta_{2}$ corrections should renormalise coupling deviations 
\begin{equation}
\zeta_{1}=\kappa_{V}-1\equiv {g_{HVV}\over g_{HVV}^\text{SM}}-1 \,,
\end{equation}
(the Lorentz structure of Eq.~\eqref{eq:oneloopc2v} is linked to the $HZZ$ interactions of Eq.~\eqref{eq:deltadef}). However, even if we introduce a bare $\zeta_1$, its renormalisation will be gauge-dependent. Therefore, $\zeta_1$ cannot directly be associated with a physical observable (in contrast to, e.g., the electromagnetic coupling $e$ and its behaviour under renormalisation group flow). 

The combination of naively weak $\kappa_{2V}$ tree-level unitarity constraints and the breaking of gauge invariance within the SM prompts us to a practical phenomenological issue when considering the analyses by ATLAS and CMS (or future experiments) within the $\kappa$ framework: {\emph{What is their phenomenological value and what is the level of $\zeta_2$ deviations that can be reasonably expected?}} 

Naturally, this question is not a sensible one in the context of the Standard Model, as highlighted by Eq.~\eqref{eq:oneloopc2v}. In concrete UV extensions, some of which we alluded to above, the answer is immediately provided by examining the $H^2V^2$ operators and their relation to the theories' input parameters; they will not be free parameters and typically be predictions of BSM electroweak input parameters that are more precisely determined from other measurements.\footnote{In the SM these couplings are rather obviously determined by the choice of precisely measured electroweak input parameters, e.g. $g_{HZZ}=eM_Z/(s_Wc_W)$, $g_{HHZZ}=e^2/(2s_W^2c_W^2)$ with $c_W^2=1-s_W^2=M_W^2/M_Z^2$.} As the analyses and measurements of \cite{ATLAS:2020jgy,CMS:2022hgz} are model-independent in the sense that they modify certain parameters whilst keeping others entirely SM-like, our task is to house these assumptions in a theoretically meaningful way. This requires an approach with which correlations are not plagued by inconsistencies displayed by Eq.~\eqref{eq:oneloopc2v} to inform the reach of $\kappa_{2V}$ deviations in the light of constraining single Higgs measurements.

%%%%%%%%%%%%%%%%%%%
\section{Relevant HEFT interactions at NLO}
\label{sec:heft}
%%%%%%%%%%%%%%%%%%%
The only theoretically plausible way to treat $H^nVV$ interactions as independent is by avoiding the physical Higgs boson being part of an electroweak doublet, which also means that Standard Model EFT (SMEFT)~\cite{Grzadkowski:2010es} is not applicable. This is possible by considering the electroweak chiral Lagrangian (or HEFT)~\cite{Susskind:1978ms,Dimopoulos:1979es,Dimopoulos:1981xc,Longhitano:1980iz,Alonso:2012px,Brivio:2013pma}, which has received considerable attention recently, including the treatment of radiative corrections~\cite{Buchalla:2013rka,Buchalla:2016sop,Gavela:2016bzc,Buchalla:2017jlu,Buchalla:2020kdh,Herrero:2020dtv,Herrero:2021iqt}. HEFT is intrinsically different from the SM, which becomes clear when radiative corrections are considered in general gauge (chiefly analysed by Herrero and Morales in~\cite{Herrero:2020dtv,Herrero:2021iqt,Herrero:2022krh}), but contains the SM as a particular parameter point choice. This enables us to achieve a theoretically consistent correlation of different Higgs channels including radiative corrections in the modern sense of renormalisability (i.e. renormalisation at a given loop order). 

As in any QFT, radiative corrections will source higher-dimensional operators, yet HEFT is non-renormalisable in the classical sense. It is therefore necessary to include all relevant operators in the renormalisation programme from the start as their renormalisation is required for a consistent final one-loop result. The $\kappa_V(\kappa_{2V})$ analysis that we would like to achieve is then related to a particular choice of couplings for the one-loop renormalised HEFT Lagrangian at a given renormalisation scale, as we will discuss below.

%%%%%%%%%%%%%%%%%%%
\subsection{HEFT Lagrangian}
\label{sec:heftlag}
%%%%%%%%%%%%%%%%%%%
Following the notations of  Refs.~\cite{Alonso:2012px,Brivio:2016fzo,Herrero:2020dtv,Herrero:2021iqt}, we consider first the leading order (chiral dimension-two~\cite{Buchalla:2013eza,Buchalla:2014eca}) HEFT Lagrangian. $W^{a}_{\mu}$ and $B_{\mu}$ are the SM gauge bosons associated with the $SU(2)_{L} \times U(1)_{Y}$ gauge symmetry. Contrary to the SM, the Higgs boson is a singlet field in HEFT and the Goldstone bosons $\pi^{a}$ are parametrised non-linearly using the matrix $U$ (a $[SU(2)_L\times SU(2)_R]/SU(2)_{R+L}$ CCWZ model)
\begin{equation}
	U(\pi^{a}) = \exp\left({i \pi^{a} \tau^{a}/v}\right)\,,
 \end{equation}
where $\tau^{a}$ are the Pauli matrices with $a= 1,2,3$. The $U$ matrix transforms under $L\in SU(2)_L, U(1)_Y\subset SU(2)_R \ni R $ as $U\to L U R^\dagger$ and is expanded as
\begin{equation}
	U(\pi^{a}) =  {\mathbbm{1}}_{2} + i \frac{\pi^{a}}{v} \tau^{a} - \frac{2G^{+}G^{-} + G^{0} G^{0}}{2 v^2}  {\mathbbm{1}}_{2} + \dots \,,
\end{equation}
where $ G^{\pm} = ( \pi^{2} \pm i \pi^{1} )/\sqrt{2}$ and $ G^{0} = -\pi^{3}$. Weak gauging of $SU(2)_L\times U(1)_Y$ is achieved through the covariant derivative
\begin{equation}
D_{\mu}U = \partial_{\mu} U + i g_{W} (W^{a}_{\mu} \tau^{a} /2) \; U -i g' U B_{\mu}\tau^{3}/2 \,.
\end{equation}
The gauge fields in the physical basis are defined as
\begin{equation}
	W^{\pm}_{\mu} =   \frac{1}{\sqrt{2}}(W^{1}_{\mu} \mp W^{2}_{\mu})\,,\quad
	\begin{pmatrix}
		Z_{\mu} \\ A_{\mu}
	\end{pmatrix}= \begin{pmatrix}
		c_{W} & s_{W} \\
		-s_{W} & c_{W}
	\end{pmatrix}
	\begin{pmatrix}
		W^{3}_{\mu} \\ B_{\mu}
	\end{pmatrix}\,.
\end{equation} 
The leading order HEFT Lagrangian relevant for this study is given as follows
\begin{subequations}
\label{eq:heftlo}
\begin{multline}
	\mathcal{L} = - \frac{1}{4} W^{a}_{\mu \nu} W^{a\mu \nu} -  \frac{1}{4} B_{\mu \nu} B^{\mu \nu} + {\cal{L}}_{\text{ferm}}+ \mathcal{L}_{\text{Yuk}} \\+ \frac{ v^2}{4} \mathcal{F}_{H} \,\text{Tr}[D_{\mu} U^{\dagger} D^{\mu} U]   
	+ \frac{1}{2} \partial_{\mu} H \partial^{\mu} H - V(H) + \mathcal{L}_{\text{GF}} + \mathcal{L}_{\text{FP}} \,,
\end{multline}
where the interactions of Higgs with gauge and Goldstone bosons are parametrised using the function $\mathcal{F}_{H}$ given as
\begin{equation}
\mathcal{F}_{H} = \Big(1+ 2(1+\zeta_{1})\frac{H}{v} + (1+\zeta_{2}) \Big(\frac{H}{v}\Big)^2 + ... \Big)\,.
\end{equation}
\end{subequations}
$\zeta_{1} $ and $\zeta_{2}$ are the new physics parameters (the choice $\zeta_{1,2}=0$ corresponds to the gauge-Higgs interactions of the SM). The ellipses denote the interaction terms with more than two Higgs bosons which are not relevant for our work. ${\cal{L}}_{\text{ferm}}$ parametrises the fermion-gauge boson interactions, which we take SM-like.
$\mathcal{L}_{\text{GF}}$ and $\mathcal{L}_{\text{FP}}$ are the gauge-fixing and Faddeev-Popov terms~\cite{Faddeev:1967fc}, respectively. $V({H})$ is the Higgs potential
\begin{equation}
V({H})=  \frac{1}{2} M_{H}^2 H^2 + \kappa_3 \frac{M_{H}^2}{2 v} H^3+ \kappa_4 \frac{M_{H}^2}{8 v^2} H^4 \,.
\end{equation}
For our analysis, we will take the potential to be SM-like, $\kappa_{3,4}=1$.
The gauge-fixing term is written in terms of gauge fixing functions as
\begin{equation}
\mathcal{L}_{\text{GF}}= - F_{+} F_{-} - \frac{1}{2} F_{Z}^2 - \frac{1}{2} F_{A}^2 \,,
\end{equation}
where the gauge fixing functions are defined as linear
\begin{equation}
	F_{\pm} = \frac{1}{\sqrt{\xi_{W}}}(\partial^{\mu} W^{\pm}_{\mu} - \xi_W M_{W} G^{\pm})\,, \hspace{0.2cm}	F_{Z} = \frac{1}{\sqrt{\xi_{Z}}}(\partial^{\mu} Z_{\mu} - \xi_Z M_{Z} G^{0})\,, \hspace{0.2cm}	F_{A} = \frac{1}{\sqrt{\xi_{A}}}(\partial^{\mu} A_{\mu})\,,
\end{equation}
with $\xi_{W}$, $\xi_{Z}$ and $\xi_{A}$ denoting the gauge-fixing parameters in the $R_{\xi}$ gauge.
Using these above gauge-fixing functions, the Faddeev-Popov part of the effective, quantised HEFT Lagrangian is then given as
\begin{equation}
	\mathcal{L}_{\text{FP}} = \sum_{i,j } {c^{i}}^{\dagger} \frac{\delta F_{i}}{\delta \alpha_{j}} c^{j}~\quad (i,j= +,-, Z,A)\,,
\end{equation}
where  $c^{j}$ are the ghost fields and $\alpha_{j}$ are the corresponding gauge transformation parameters which are similar to that of SM. 
We parametrise the couplings of Higgs boson with fermions via
\begin{equation}
	 \mathcal{L}_{\text{Yuk}} = - \frac{v}{\sqrt{2}} \begin{pmatrix}
	 	\bar{u}^i_{L}  & \bar{d}^i_{L}
	 \end{pmatrix} U \Big(1+ c\,\frac{h}{v}+ ...\Big) \begin{pmatrix}
	 y^{u}_{ij} u^{j}_{R}  \\ y^{d}_{ij} d^{j}_{R} 
 \end{pmatrix} + \text{h.c.}\,,
 \end{equation} 
where $y^{u}_{ij}$ and $y^{d}_{ij}$ are the Yukawa couplings. In this study, we will focus on the case $c=1$, which after diagonalisation, directly links the Yukawa couplings to the fermion mass eigenstates
\begin{equation}
y^u_{33} ={\sqrt{2} m_t\over v}\,, \quad y^d_{33} = {} {\sqrt{2}m_b\over v}\,,
\end{equation}
with top and bottom masses $m_{t,b}$, respectively. We will neglect the light quark flavour and lepton masses throughout this work since these do not impact our phenomenological findings.

When considering radiative corrections in the effective theory described by Eq.~\eqref{eq:heftlo}, ultraviolet divergences will source new operator structures at one-loop level. This requires the introduction of bare terms to the Lagrangian {\it ab initio} for a consistent, one-loop renormalisation programme~\cite{Herrero:2021iqt,Herrero:2022krh}. The terms relevant at NLO one-loop level (chiral dimension 4) are
\begin{equation}
	\label{eq:chir4op}
	{\cal{L}}_4 = \sum_i a_i {\cal{O}}_i\,,
\end{equation}
with operators that require UV regularisation listed in Tab.~\ref{tab:operators}.

%%%%%%%%%%%%%%%%%%%%%%%%%
\begin{table}[t!]
	\centering
	\renewcommand{\arraystretch}{1.7}
  \small{	\begin{tabular}{|c|c|}
		\hline 
		$\mathcal{O}_{0}$&
		$a_{0} (M_Z^2 - M_W^2) \text{Tr}\Big[U \tau^3 U^{\dagger} \boldsymbol{\mathcal{V}}_{\mu} \Big] \text{Tr}\Big[U \tau^3 U^{\dagger} \boldsymbol{\mathcal{V}}_{\mu} \Big]$  \\
		\hline
		$\mathcal{O}_{1}$ &
		$a_{1} \; g' g_{W} \text{Tr}\Big[U B_{\mu\nu} \frac{\tau^3}{2} U^{\dagger} W^{a}_{\mu\nu}\frac{\tau^{a}}{2} \Big]$  \\
		\hline
	   $\mathcal{O}_{HBB}$ 
	  &$-a_{HBB}  \; g'^2\frac{H}{v} \text{Tr}\Big[ B_{\mu \nu} B^{\mu \nu}\Big]$ \\
	  \hline
		$\mathcal{O}_{HWW}$ 
		&$ -a_{HWW}\;g_{W}^2 \frac{H}{v} \text{Tr}\Big[ W^{a}_{\mu \nu} W^{a\mu \nu}\Big]$ \\
		\hline
		$\mathcal{O}_{\square \boldsymbol{\mathcal{V}} \boldsymbol{\mathcal{V}} }$ 
		& $a_{\square \boldsymbol{\mathcal{V}} \boldsymbol{\mathcal{V}}}  \frac{\square H}{v} \text{Tr}\Big[ \boldsymbol{\mathcal{V}}_{\mu} \boldsymbol{\mathcal{V}}^{\mu}\Big]$ \\
		\hline
		$\mathcal{O}_{H0}$&
		$a_{H0} (M_Z^2 - M_W^2) \frac{H}{v} \text{Tr}\Big[U \tau^3 U^{\dagger} \boldsymbol{\mathcal{V}}_{\mu} \Big] \text{Tr}\Big[U \tau^3 U^{\dagger} \boldsymbol{\mathcal{V}}_{\mu} \Big]$  \\
		\hline
		$\mathcal{O}_{H1}$&
		$a_{H1} \; g' g_{W}\frac{H}{v} \text{Tr}\Big[U B_{\mu\nu} \frac{\tau^3}{2} U^{\dagger} W^{a}_{\mu\nu}\frac{\tau^{a}}{2}\Big]$  \\
		\hline
		$\mathcal{O}_{H11}$&
		$a_{H11} \frac{ H}{v} \text{Tr}\Big[ \mathcal{D}_{\mu} \boldsymbol{\mathcal{V}}^{\mu} \mathcal{D}_{\nu}\boldsymbol{\mathcal{V}}^{\nu}\Big]$ \\
		\hline
		$\mathcal{O}_{d1}$ 
		& $ i a_{d1} \; g' \frac{\partial^{\nu} H}{v} \text{Tr}\Big[ U B_{\mu \nu} \frac{\tau^3}{2} U^{\dagger} \boldsymbol{\mathcal{V}}^{\mu}\Big]$ \\
		\hline
		$\mathcal{O}_{d2}$ 
		& $ i a_{d2} \; g_{W} \frac{\partial^{\nu} H}{v} \text{Tr}\Big[ W^{a}_{\mu \nu}\frac{\tau^{a}}{2} \boldsymbol{\mathcal{V}}^{\mu}\Big]$ \\
		\hline
		$\mathcal{O}_{d3}$ 
		& $ a_{d3}  \frac{\partial^{\nu} H}{v} \text{Tr}\Big[  \boldsymbol{\mathcal{V}}^{\mu} \mathcal{D}_{\mu}\boldsymbol{\mathcal{V}}^{\mu}\Big]$ \\
		\hline
		$\mathcal{O}_{\square \square}$ 
		& $a_{\square \square }  \frac{\square H \square H}{v}$ \\
		\hline
	\end{tabular}}
	\caption{HEFT operators $\mathcal{O}_{i}$ required for the cancellations of divergences, $a_{i}$ are the corresponding HEFT coefficients. $\boldsymbol{\mathcal{V}}_{\mu}= (D_{\mu}U)U^{\dagger}$ and $\mathcal{D}_{\mu}\boldsymbol{\mathcal{V}}^{\mu}= \partial_{\mu}\boldsymbol{\mathcal{V}}^{\mu} + i [g_{W}W^{a}_{\mu}\frac{\tau^{a}}{2},\boldsymbol{\mathcal{V}}^{\mu}] $.} 
	\label{tab:operators}
\end{table}  
%%%%%%%%%%%%%%%%%%%%%%%%%

%%%%%%%%%%%%%%%%%%%%%%%%%
\subsection{Amplitudes and Renormalisation}
\label{sec:ren}
%%%%%%%%%%%%%%%%%%%%%%%%%
Throughout, we will work in dimensional regularisation $D=4-2\ep$~\cite{tHooft:1972tcz} and adopt the on-shell scheme for propagating degrees of freedom~\cite{Denner:1991kt}. Expanding out the non-linear sigma model of Eq.~\eqref{eq:heftlo}, we understand the electroweak vev as a derived quantity
\begin{equation}
v = {2 M_W s_W \over e}\,,
\end{equation}
with 
\begin{equation}
s^2_W = 1 -c_W^2 = 1 - {M_W^2\over M_Z^2}\,,
\end{equation}
following Sirlin~\cite{Sirlin:1980nh,Marciano:1980pb}.
These can be parametrically fixed through choices of $e=\sqrt{4\pi \alpha} ,M_W$ and $M_Z$. As we identify these quantities as the input parameters for the gauge part of the Lagrangian (coupling constants of $SU(2)_L\times U(1)_Y$ are parametrised as $g_W=e/s_W, g'=e/c_W$, respectively), we have to supply counter terms $\dze,\dmw,\dmz$ for the bare Lagrangian quantities (we indicate unrenormalised quantities with a subscript~`0' in Eq.~\eqref{eq:heftlo},~\eqref{eq:chir4op} after ${\cal{L}}\to {\cal{L}}_0$)
\begin{equation}
\begin{split}
e_0 & = Z_e e = (1+\delta Z_e) e\,,\\
M_{0,W}^2 &= M_W^2 + \dmw\,, \\
M_{0,Z}^2 & =M_Z^2 + \dmz\,.
\end{split}
\end{equation}
The renormalisation of the Weinberg angle 
\begin{equation}
s_{W,0}=s_{W}+\dsw\,,\quad c_{W,0}=c_{W}+\dcw \,,
\end{equation}
is then given by
\begin{equation}
\label{eq:weinberg}
\dsw = {c_W^2 \over 2 s_W} \left( {\dmz\over M_Z^2 } - {\dmw \over M_W^2} \right)\,,\quad \dcw=c_W - {s_W\over c_W} \dsw\,.
\end{equation}
The field renormalisations of the bare Lagrangian quantities are
\begin{equation}
\begin{split}
W^{\pm}_{0,\mu}&=\sqrt{Z_W} W^\pm_\mu = (1+\delta Z_W/2)W^\pm_\mu\,,\\
H_0 &= \sqrt{Z_H} H = (1+\delta Z_H/2) H\,,
\end{split}
\end{equation}
and
\begin{equation}
\left(\begin{matrix} A_0 \\ Z_0 \end{matrix} \right)_\mu
= \left(\begin{matrix}  \sqrt{Z_{AA}} & \sqrt{Z_{AZ}} \\ \sqrt{Z_{ZA}} & \sqrt{Z_{ZZ}} \end{matrix} \right) \left(\begin{matrix} A \\ Z \end{matrix} \right)_\mu
=\left(\begin{matrix} 1 + \dzaa & \dzaz \\ \dzza & 1 +\dzzz \end{matrix} \right) \left(\begin{matrix} A \\ Z \end{matrix} \right)_\mu \;,
\end{equation}
for the photon-$Z$ sector. Tadpoles are renormalised by requiring that they fully cancel against the tadpole counter term. They are treated as part of the parameter renormalisation~\cite{Denner:1991kt,Denner:2018opp,Dudenas:2020ggt} so that the tadpole counter term only enters the unphysical Goldstone sector and we can ignore associated contributions in the following. The HEFT coefficients are renormalised through
\begin{equation}
\label{eq:wilcoren}
\zeta_{0,i} = \zeta_{i} + \dz_i\,, \quad a_{0,i} = a_{i} + \da_i\,.
\end{equation}
We choose on-shell (OS) renormalisation conditions for the gauge and Higgs sectors, i.e. the wave function renormalisations are chosen to yield unit residues for the propagators at the physical masses $M_W^2,M_Z^2,M_H^2$; we remove $Z-\gamma$ mixing for on-shell kinematics, $\sigaz(q^2=0) = \sigza(q^2=M_Z^2) = 0$ (where $\Sigma^T$ denotes the transverse part of the self-energy in case of the vector bosons). Alongside on-shell renormalisation in the fermion-sector $f$ (the form of which is not relevant for our work), this reduces the renormalisation condition for the electromagnetic coupling constant to the condition of the amputated three-point function $\Gamma_{f\bar f \gamma}^\mu = -ie\gamma^\mu$ for on-shell legs and zero momentum transfer. The renormalisation constants can then be worked out as functions of the self-energies and their derivatives and are tabled in, e.g.,~\cite{Denner:1991kt,Denner:2019vbn}. From this follows in particular
\begin{equation}
\delta Z_e = -{1\over 2} \left( \dzaa + {s_W\over c_W} \dzza \right)\,.
\end{equation}
For the coefficients of Eq.~\eqref{eq:wilcoren}, we choose the \MSb~scheme.

Expanding the Lagrangian of Eq.~\eqref{eq:heftlo} into a format compatible with {\sc{FeynRules}}~\cite{Christensen:2008py,Alloul:2013bka}, we perform our calculations using {\sc{FeynArts}}, {\sc{FormCalc}}, and {\sc{LoopTools}}~\cite{Hahn:1998yk,Hahn:2000jm,Hahn:2000kx,Hahn:2006qw} alongside {\sc{PackageX}}~\cite{Patel:2016fam} for numerical evaluations and analytical cross-checks. Calculating in general gauges $\sim \xi_A,\xi_W,\xi_Z$, we have verified gauge independence at the level of divergent parts as well as finite parts (see below). Further checks include comparisons against existing results: In particular Ref.~\cite{Herrero:2021iqt} provides a comprehensive overview of the HEFT Lagrangian at one-loop. Given that we find agreement with the results quoted there, we limit ourselves to a summary of the findings relevant for our analysis in Sec.~\ref{sec:constraints}.

%%%%%%%%%%%%%%%%%%%%%%
\subsection*{$H\to WW^\ast $ amplitude renormalisation}
\begin{figure}[!t]
\centering \includegraphics[width=12cm]{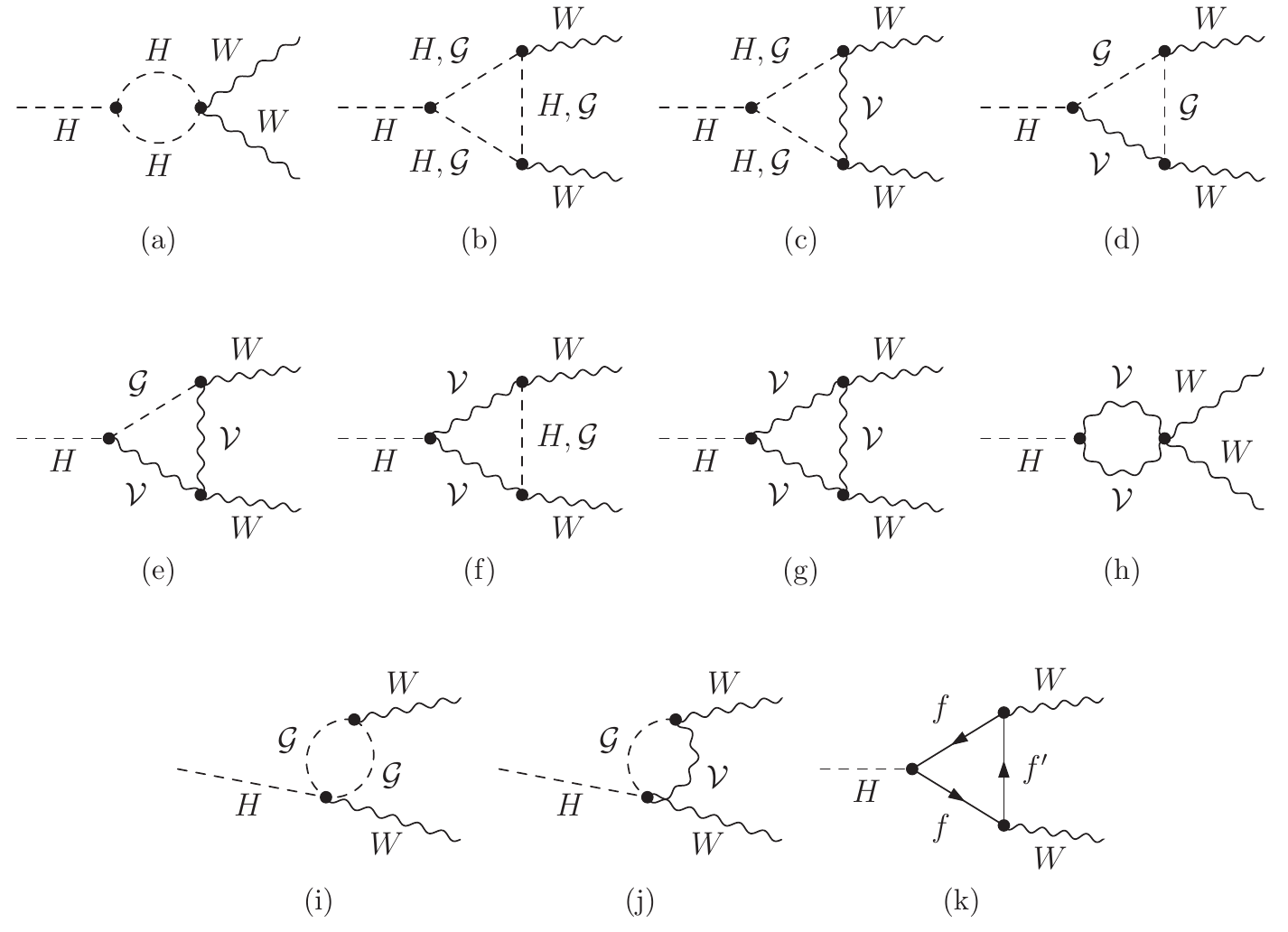}
\caption{\label{fig:wdiags} Representative Feynman diagram topologies contributing to $H\to WW$. We use abbreviations ${\cal{V}}=\{W,Z,A\}$ for the vector bosons and ${\cal{G}}=\{G^0,G^+\}$ for the would-be Goldstone bosons. Note that we only consider the third family of fermions $f=\{t,b\}$ as massive for the purpose of this work.}
\end{figure}

%%%%%%%%%%%%%%%%%%%%%%
The (off-shell) $H(q)W^\mu(k_1)W^\nu(k_2)$ one-loop three-point function is pictorially represented by the Feynman diagrams of Fig.~\ref{fig:wdiags}. We limit ourselves to transverse polarisations. The counter term vertex (in a convenient phase convention) follows from the procedure described in Sec.~\ref{sec:ren}, see also appendix~\ref{sec:uvdiv},
 \begin{multline}
\parbox{2.5cm}{\includegraphics[width=2.5cm]{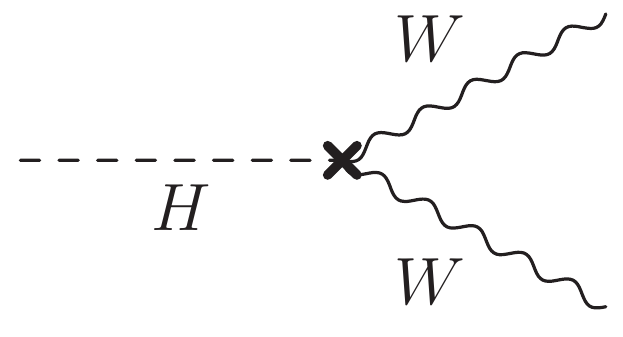}}
= { e M_W(1+\zeta_1) \over s_W }  \left( \delta Z_e + {\delta M_W^2 \over 2 M_W^2} - {\delta s_W\over s_W} + {\delta \zeta_1\over 1+ \zeta_1} + {\delta Z_H \over 2} + {\delta Z_W} \right) g^{\mu\nu} \\
+ \delta
\left[\parbox{2.5cm}{\includegraphics[width=2.5cm]{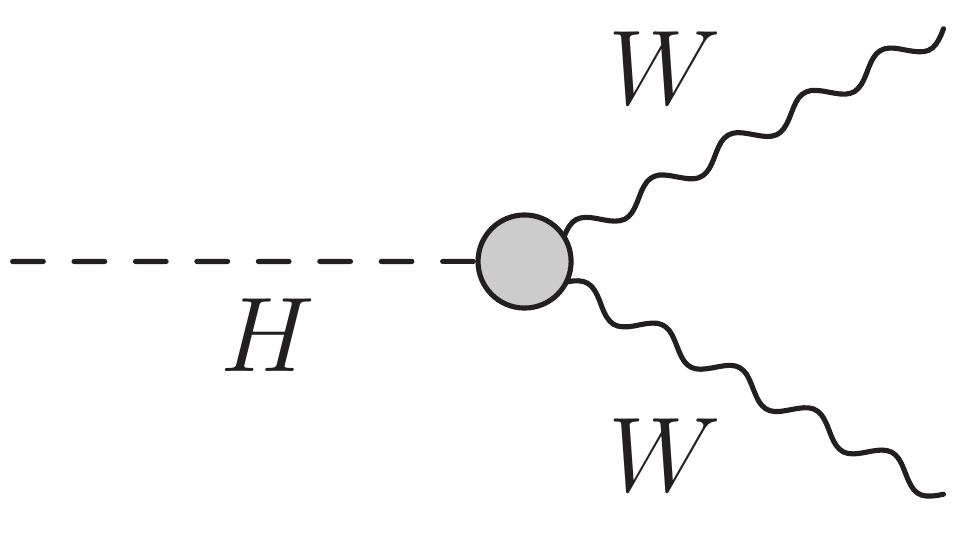}}\right]\,,
\end{multline}
with $\delta[...]$ summarising the renormalisation contributions from Eq.~\eqref{eq:chir4op}.
Contributions proportional to the matrix element $\sim \left\langle HW^{\mu \nu}W_{\mu\nu}\right\rangle$ are UV-finite, which imposes a cancellation\footnote{The notation $|_\Delta$ refers to the part proportional to the \MSb{} factor of Eq.~\eqref{eq:deltadef}, i.e. $\delta a|_\Delta = \delta a/\DeltaUV$.}
\begin{equation}
\label{eq:ahwwren}
\delta a_{HWW}\big|_\Delta = - {1\over 2} \delta a_{d2}\big|_\Delta = {1\over 192\pi^2} (1+\zeta_1) (\zeta_1^2+2\zeta_1-\zeta_2)\,.
\end{equation}
The remaining singularity $\sim q^2 \left\langle HW^{\mu }W_{\mu}\right\rangle$ then allows us to read off 
\begin{equation}
\label{eq:aboxren}
\delta a_{\Box{\boldsymbol{\cal{V}}}{\boldsymbol{\cal{V}}}}\big|_\Delta = - {1\over 64\pi^2} (1+\zeta_1)\left(2+(1+\zeta_1)^2\right)\,.
\end{equation}
The precise form of these additional terms is not relevant for the analysis of Sec.~\ref{sec:constraints}, but we quote these here to highlight the agreement with~\cite{Herrero:2021iqt}. The singularity related to the bare momentum-independent $HWW$ vertex, on the other hand, fixes
\begin{multline}
\label{eq:zeta1}
\delta \zeta_1\big|_\Delta = {3 \alpha \over 32\pi s_W^2 M_W^2} \big ( 4 (m_b^2+m_t^2) \zeta_1 + M_Z^2 (3s_w^2-4) \zeta_1 (1+\zeta_1)(2+\zeta_1) \\+M_H^2 (1-\kappa_3+\zeta_1) (\zeta_1 (2+\zeta_1) - \zeta_2) + 2M_W^2(1+\zeta_1)\zeta_2\big)\,,
\end{multline}
after entering the renormalisation constants of appendix~\ref{sec:uvdiv}.

%%%%%%%%%%%%%%%%%%%%%%
%%%%%%%%%%%%%%%%%%%%%%

%%%%%%%%%%%%%%%%%%%%%%
\begin{figure}[!t]
\centering \includegraphics[width=12cm]{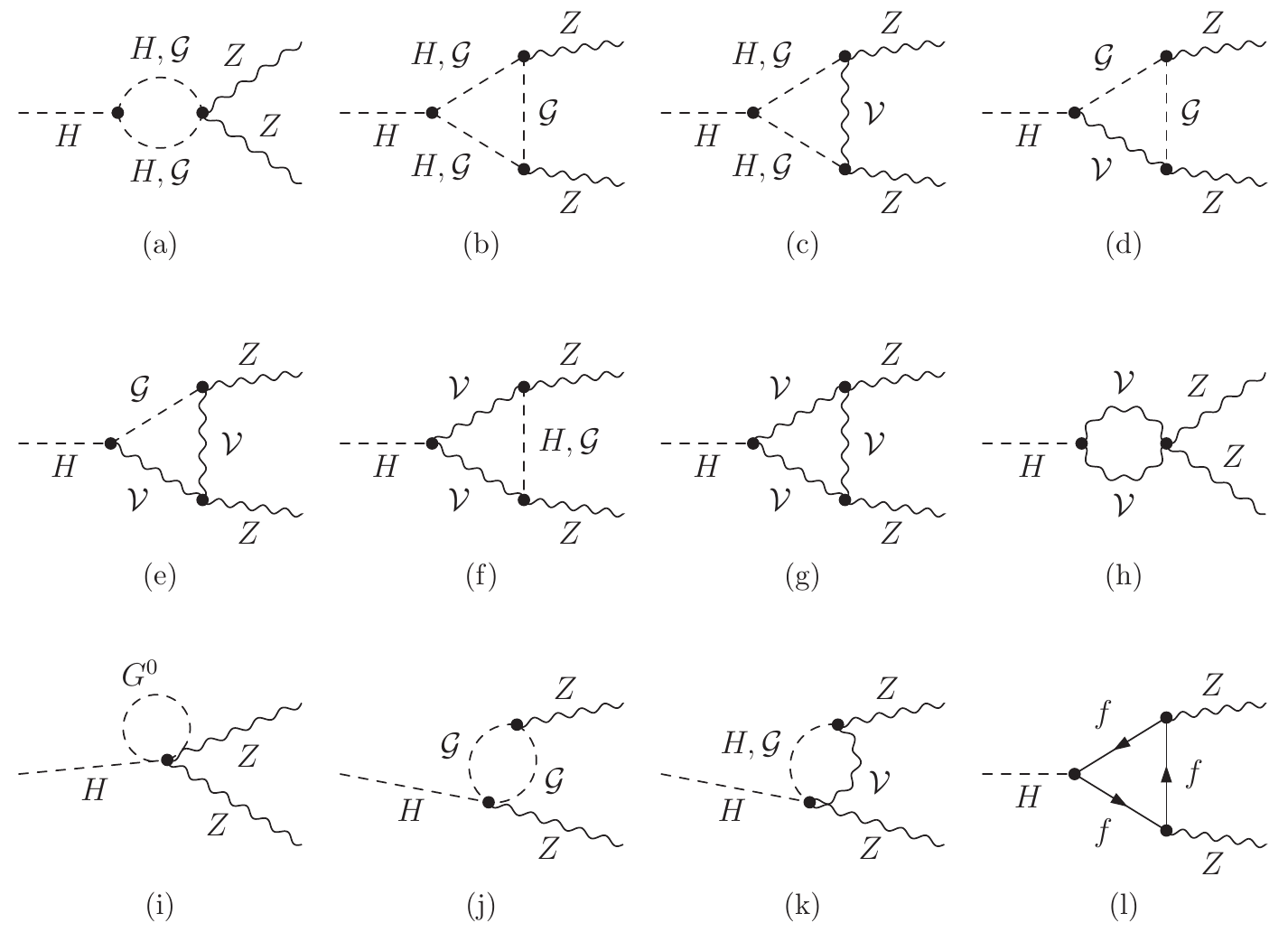}
\caption{\label{fig:zdiags} Representative Feynman diagram topologies contributing to $H\to ZZ$, similar to Fig.~\ref{fig:wdiags}. 
Note that due to the structure of the HEFT Lagrangian, we obtain 5-point interactions in the $H\to ZZ$ channel (i).}
\end{figure}
%%%%%%%%%%%%%%%%%%%%%%
%%%%%%%%%%%%%%%%%%%%%%
\subsection*{$H\to ZZ^\ast $ amplitude renormalisation}
%%%%%%%%%%%%%%%%%%%%%%
Similar to the discussion above, the (off-shell) $H(q)Z(k_1)Z(k_2)$ one-loop three-point function is summarised in Fig.~\ref{fig:zdiags}. The counter term vertex is
 \begin{multline}
\parbox{2.5cm}{\includegraphics[width=2.5cm]{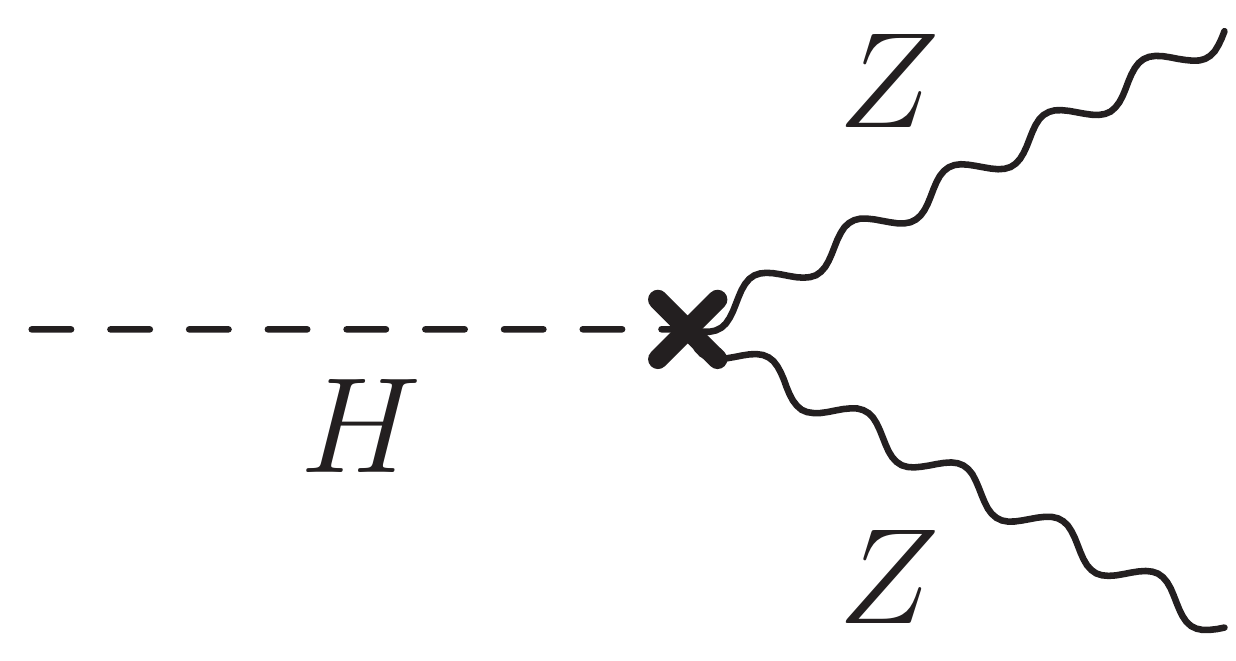}}
= { e M_Z  (1+\zeta_1) \over s_W c_W } \left( \delta Z_e + {\delta M_Z^2 \over 2 M_Z^2} - {\delta s_W\over s_W}  \right. \\  \left.- {\delta c_W\over c_W} + {\delta \zeta_1\over 1+ \zeta_1} + {\delta Z_H \over 2} + {\dzzz} \right)g^{\mu\nu} + \delta
\left[\parbox{2.5cm}{\includegraphics[width=2.5cm]{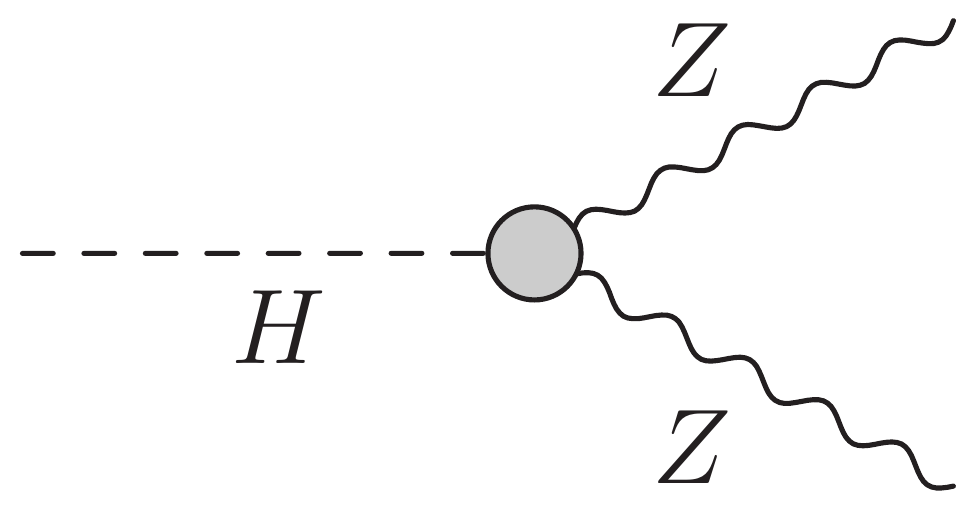}}\right]\,.
\end{multline}
and there are similar cancellations for the counter terms $\delta[\dots]$ leading to Eq.~\eqref{eq:ahwwren}, which are not relevant for the analysis of Sec.~\ref{sec:constraints}. Entering Eq.~\eqref{eq:zeta1} then fixes the remaining
\begin{equation}
\delta a_{H0}\big|_\Delta = {3\over 64\pi^2} (1+\zeta_1)\zeta_2\,,
\end{equation}
when including the renormalisation constants of appendix~\ref{sec:uvdiv}.

%%%%%%%%%%%%%%%%%%%%%%
\subsection*{$H\to \gamma Z$ amplitude renormalisation}
%%%%%%%%%%%%%%%%%%%%%%
The $H\to \gamma Z$ counter term amplitude is given by
\begin{equation}
\parbox{2.5cm}{\includegraphics[width=2.5cm]{zzct.pdf}} = {e M_W \over 2 c_W^2 s_W}  \dzza  + \delta \left[\parbox{2.5cm}{\includegraphics[width=2.5cm]{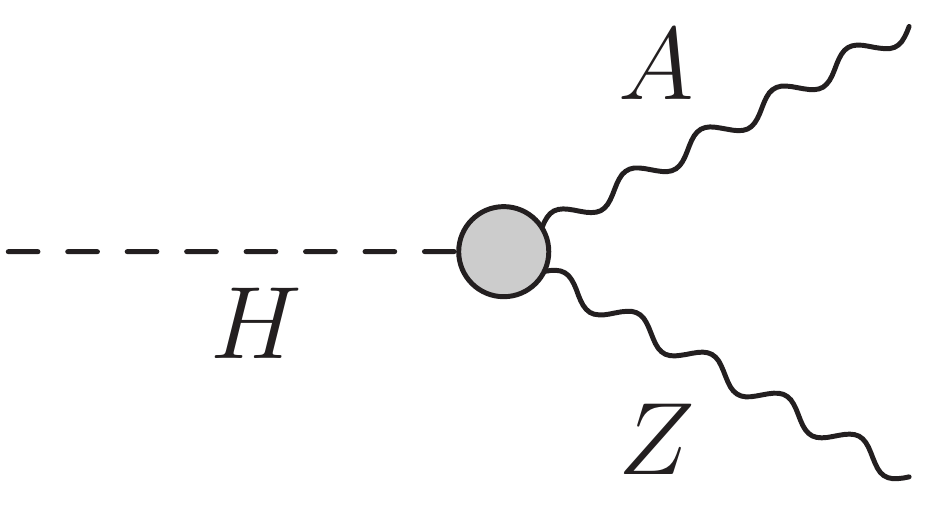}}\right]\,,
\end{equation}
which renders the off-shell amplitude UV finite (we do not show the Feynman diagram topologies as they are similar to Fig.~\ref{fig:zdiags}).
Again the terms $\delta[\dots]$ enforce non-trivial identities among the one-loop counter terms of the operators of Eq.~\eqref{eq:chir4op}, which are again not relevant for our $\kappa_{2V}-\kappa_V$ analysis. Entering the wave-function renormalisation constant of appendix~\ref{sec:uvdiv}, we obtain a UV-finite on-shell $H\to \gamma Z$ amplitude which is manifestly gauge-independent. Our result agrees with the findings of~\cite{Herrero:2020dtv} and the unitary gauge calculation of~\cite{deBlas:2018tjm}.

%%%%%%%%%%%%%%%%%%%%%%
\subsection*{$H\to \gamma \gamma$ amplitude renormalisation}
%%%%%%%%%%%%%%%%%%%%%%
The $H\to \gamma \gamma$ amplitude has been first computed in general $R_\xi$ gauge in~\cite{Herrero:2020dtv} after unitary gauge results have been used in~\cite{deBlas:2018tjm}. The $H\to \gamma\gamma$ is free of singularities (see also~\cite{Weinberg:1971fb,Marciano:2011gm}) and therefore only imposes one-loop counter term relations among the coefficients of Eq.~\eqref{eq:chir4op}. Again these are not relevant for the analysis below, but we have cross-checked our findings against the existing results in the literature.
\bigskip

%%%%%%%%%%%%%%%%%%%%%%%%%
\subsection{Decay widths}
%%%%%%%%%%%%%%%%%%%%%%%%%
The amplitudes $H\to \gamma\gamma, \gamma Z$ of the previous section can be converted into decay widths $\Gamma(H\to \gamma\gamma), \Gamma(H\to \gamma Z)$ straightforwardly; they are only sensitive to $\zeta_1$. For the cases of $H\to ZZ^\ast$ and $H\to WW^\ast$ we follow the SMEFT analysis of Refs.~\cite{Dawson:2018liq,Dawson:2018pyl}. To this end, we write the decay width $H(p_H) \to f(p_1) f(p_2) V(p_3)$ as
\begin{equation}
	\label{eq:offshellhvv}
	\Gamma(H\to VV^\ast) = \int_{0}^{(M_H - M_V)^2} \hbox{d} m_{12}^2 \int_{m_{23,\text{l}}^2}^{m_{23,\text{u}}^2} \hbox{d} m_{23}^2\, \dfrac{{{\cal{M}}}^2}{ (2\pi)^3\, 32 \,M_H^3}\,,
\end{equation}
with $m_{ij}^2 = (p_i + p_j)^2$. The invariant masses satisfy the relation 
\begin{equation}
m_{23}^2 + m_{23}^2 + m_{13}^2 = M_H^2 + M_V^2\,.
\end{equation}
With the K\"all\'en lambda function denoted as 
\begin{equation}
\lambda(m_{12},M_H,M_V) = m_{12}^4 - 2 m_{12}^2 (M_H^2 + M_V^2) + (M_H^2 - M_V^2)^2\,, 
\end{equation}
the limits of the $m_{23}^2$ integration are 
\begin{equation}
m_{23,\text{l},\text{u}}^2={1\over 2}\left(M_H^2 + M_V^2 - m_{12}^2 \mp \sqrt{\lambda(m_{12},M_H,M_V)}\right)\,.
\end{equation}
The amplitudes are constructed by considering the Eq.~\eqref{eq:offshellhvv} expanded to one-loop level
\begin{equation}
{\cal{M}}^2 =  |{\cal{M}}_{\text{LO}}|^2 + 2\,\text{Re}\left\{ {\cal{M}}_{\text{LO}} {\cal{M}}_{\text{1-loop}}^\ast \right\} \,,
\end{equation} 
and both the leading order (LO) and one-loop parts can be written as
\begin{equation}
{\cal{M}} = {\cal{M}}^\mu \epsilon^\ast_{\mu} = {\cal{M}}_{HVV}^{\mu \nu} \Delta_{VV}^{\nu \rho} {{\cal{M}}_{f f V}^{\rho}} \epsilon^\ast_{\mu}\,,
\end{equation}
where the polarisation vector of the on-shell boson is denoted with $\epsilon_\mu$. ${\cal{M}}_{ffV}^\rho$ is the $ffV$ vertex arising from $V$-current and includes the on-shell spinors, while $\Delta_{VV}^{\nu \rho}$ is the vector boson propagator. We do not include higher order corrections for these parts and only consider them for the $HVV$ part of the amplitude, ${\cal{M}}_{HVV}^{\mu \nu}$. In the case of $H\to ZZ^\ast$ we also do not include the $H\to \gamma Z$ contribution (similar to~\cite{Dawson:2018pyl}). 

The renormalised $HVV$ vertices described above can be decomposed into three different Lorentz structures using the metric tensor $g^{\mu\nu}$ and the Levi-Civita antisymmetric tensor $\varepsilon^{\mu \nu \rho \sigma}$ as
\begin{equation}
	{\cal{M}}_{HVV}^{\mu \nu} = (F_1^\text{LO} + F_1^\text{NLO}) g^{\mu \nu} + F_2^{\text{NLO}} (p_1 + p_2)^\mu p_3^\nu + F_3^{\text{NLO}} \tensor{\varepsilon}{^\mu^\nu_\rho_\sigma} p_H^\rho (p_1 + p_2)^\sigma\;.
\end{equation}
The last two terms only contribute at one-loop order for the SM and HEFT for the choices detailed in Sec.~\ref{sec:ren}. $F_3^{\text{NLO}}$ is zero for the $H \to f \bar f Z$ channel, while the longitudinal-like second term does not contribute as we consider massless fermions. The non-trivial form factors are directly related to the one-loop calculation detailed above. By considering the unpolarised amplitude and performing the integrations of Eq.~\eqref{eq:offshellhvv} we verify that our calculations reproduce the expressions supplied by Refs.~\cite{Dawson:2018dcd,Dawson:2018liq,Dawson:2018pyl} for the SM case at LO (see also~\cite{Rizzo:1980gz,Keung:1984hn}). For HEFT with NLO corrections, the integrations are performed numerically to obtain the decay widths $\Gamma(H \to W W^\ast)$ and $\Gamma(H \to Z Z^\ast)$ and subsequently the relevant $\kappa_V=1+\zeta_1$ for a scale choice $\mu_R=M_H$. For an efficient evaluation of the amplitudes, we furthermore choose Feynman Gauge $\xi_W=\xi_Z=\xi_A=1$ and rescale our results to the values reported by the Higgs Cross Section Working Group~\cite{LHCHiggsCrossSectionWorkingGroup:2011wcg} for the SM point $\zeta_1=\zeta_2=0$ for ease of comparison.

\subsection{Comments on $\kappa_{2V}-\kappa_{V}$ interpretations}
The cancellation of gauge fixing parameters $\xi_{W,Z,A}$ in the amplitude calculations detailed above is highly non-trivial. Indeed, individual Feynman diagrams of Figs.~\ref{fig:wdiags} and~\ref{fig:zdiags} contain singularities $\sim \xi_{W,Z}^2,\xi_{W,Z,A}$, which cancel when we sum over the contributing diagrams. For instance, the vanishing $\xi^2_{W,Z}\DeltaUV$ dependence of $H(q)\to W(k_1) W(k_2)$ results from a cancellation between the gauge boson insertions of diagrams Fig.~\ref{fig:wdiags} (g) and (h) like in the SM. A zero $\sim \xi_{W,Z} \DeltaUV$ coefficient results from cancellations between diagrams Fig.~\ref{fig:wdiags} (b), (c), (d), (i) and (j), and $\sim \xi_{A} \DeltaUV$ is due to destructive interference between (c), (g), (j). Similar cancellations take place for $H\to ZZ$ (there is dependence on $\xi_A$).

As can be seen from Eqs.~\eqref{eq:ahwwren} and~\eqref{eq:aboxren} (see also appendix~\ref{sec:uvdiv}), terms of Eq.~\eqref{eq:chir4op} are sourced even for $\zeta_1=\zeta_2=0$. Although we must include all relevant HEFT coefficient renormalisation factors in the calculation, we can obtain a consistent $\kappa_{2V}-\kappa_{V}$ correlation at our scale choice $\mu_R=M_H$ by projecting all amplitudes to $a_{i}=0$ after renormalisation. Concretely this means that we will limit our attention to $\zeta_{1,2}$ in the following and choose all other $a_i(M_H)=0$ as \MSb{} parameters. In particular, this means that operator matrix elements $\left \langle HV^{\mu\nu}V_{\mu\nu} \right \rangle  (V=A,W,Z)$ are absent, which would otherwise impact the Higgs boson production and decay phenomenology. These should be included in a comprehensive fit of the HEFT Lagrangian that we do not attempt here. 

%%%%%%%%%%%%%%%%%%%
\begin{table}[b!]
	\centering
	\renewcommand{\arraystretch}{1.8}
	\begin{tabular}{|c|c|c|cccc|}
		\hline 
		Parameters 		&  ATLAS Run 2 data~\cite{ATLAS:2022vkf}	& HL-LHC uncertainties 	& \multicolumn{4}{c|}{Correlation Matrix~\cite{ATLAS:2022vkf}} \\ 
		\hline 
		$\kappa_Z$ 			& $0.99^{+0.06}_{-0.06}$  &$\pm0.012$& $1$ & $0.40$ & $0.44$ & $0.09$\\
		$\kappa_W$ 			& $1.05^{+0.06}_{-0.06}$ & $\pm0.013$&  & $1$& $0.47$ &$0.08$   \\
		$\kappa_{\gamma}$ 	& $1.01^{+0.06}_{-0.06}$ & $\pm0.013$ & & & $1$ & $0.12$ \\
		$\kappa_{Z\gamma}$ 			& $1.38^{+0.31}_{-0.37}$ &$\pm0.073$ 	& & & & 1  \\ 
		\hline
	\end{tabular}
	\caption{Constraints used for Fig.~\ref{fig:deltas}, based on~\cite{ATLAS:2022vkf}. Columns~2 and 4 show the  ATLAS Run-2 results along with the correlation matrix. In column 3, the HL-LHC projected uncertainties are listed. }
	\label{tab:exp_kappa}
\end{table}
%%%%%%%%%%%%%%%%%%%

%%%%%%%%%%%%%%%%%%%
\section{$\kappa_{V}-\kappa_{2V}$ correlations from Higgs data}
\label{sec:constraints}
\begin{figure}[!t]
	\centering 
	\subfloat[]{\includegraphics[width=7.4cm]{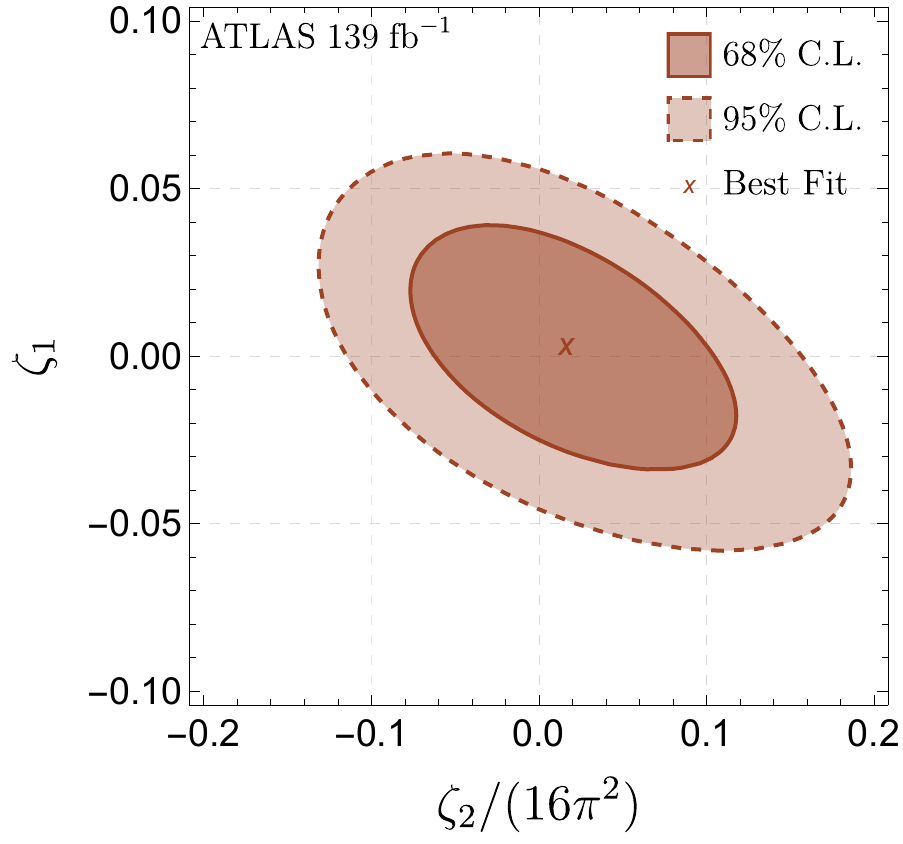}\label{fig:deltachi2_kappa} }~~
	\centering  \subfloat[]
	{\includegraphics[width=7.4cm]{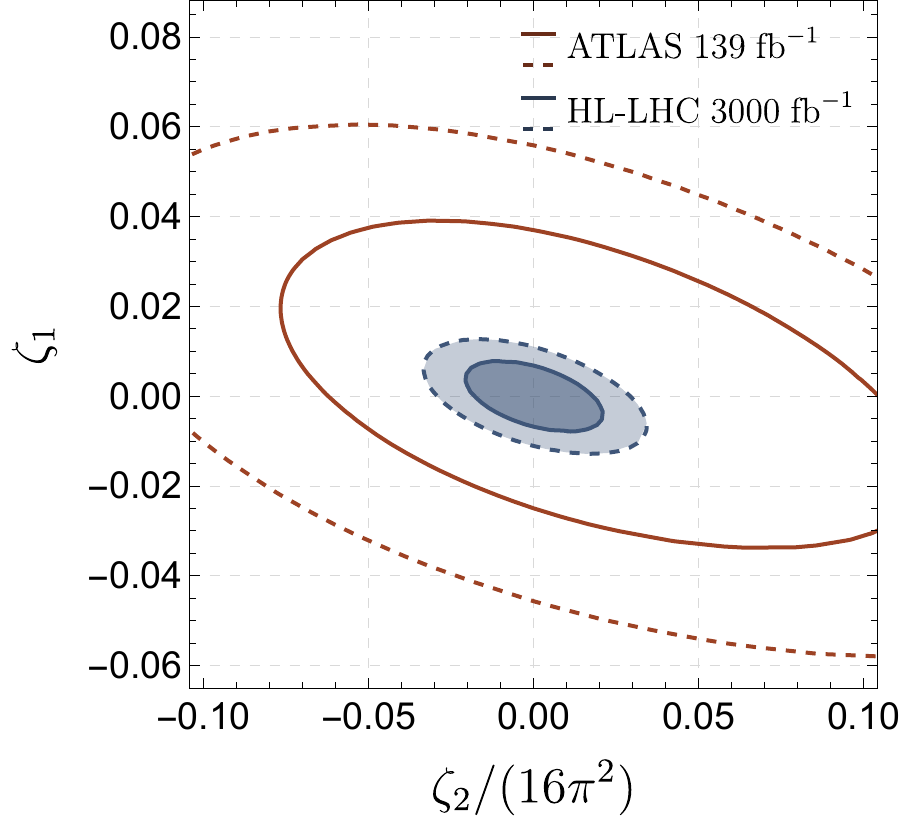}\label{fig:deltachi2_kappa_proj}} 
	\caption{The $\zeta_{1}$ vs. $ \zeta_{2}$ parameter space obtained from the $\chi^2$ analysis of the $\kappa$ parameters from ATLAS Run-2 data~\cite{ATLAS:2022vkf}. The solid (dashed) contour denotes the  $\Delta\chi^2=$ 2.28 ($\Delta\chi^2=$ 5.99) which corresponds to confidence level of $68\%$ ($95\%$). Fig.~\ref{fig:deltachi2_kappa_proj} shows the similar regions in blue obtained with the HL-LHC projected data.  \label{fig:deltas}  }
\end{figure}

%%%%%%%%%%%%%%%%%%%
Equipped with the findings of the previous sections, we are now ready to turn to results. For the current status of the Higgs programme, we construct a $\chi^2$ test from the $\kappa$ results provided in Ref.~\cite{ATLAS:2022vkf}. The $\chi^2$ statistic defined from the $\kappa$ data at an integrated luminosity of 139 $\text{fb}^{-1}$ is given by
\begin{equation}\label{eq:chi2}
	\chi^2(\zeta_{1},\zeta_2) = \sum_{i,j = 1 }^{\text{data}}\big(\kappa_{i, \text{exp}}- \kappa_{i,\text{th}}(\zeta_{1},\zeta_2)\big) (V_{ij})^{-1} \big(\kappa_{j, \text{exp}}- \kappa_{j,\text{th}}(\zeta_{1},\zeta_2)\big)\,,
\end{equation}
where $\kappa_{i, \text{exp}}$ are the central values and $V_{ij}$ is the covariance matrix, the elements of which are given as $\rho_{ij} \sigma_{i} \sigma_{j}$ with $\sigma_{i}$ as uncertainties  and   $\rho_{ij}$ as correlations.  The details of the experimental  data are provided in Tab.~\ref{tab:exp_kappa}.
Using the four $\kappa$ parameters detailed above in Eq.~\eqref{eq:chi2}, we perform a $\chi^2$ fit for $\zeta_{1}$ and $\zeta_{2}$.\footnote{For 68$\%$ (95$\%$) confidence level (C.L.), $\Delta\chi^2 = 2.28$ ($\Delta\chi^2 = 5.99$) for two degrees of freedom (d.o.f).}
The allowed regions are shown in Fig.~\ref{fig:deltachi2_kappa}. 

%%%%%%%%%%%%%%%%%%%
%%%%%%%%%%%%%%%%%%%

To gauge the improvements that can be expected at the high luminosity (HL) LHC phase, we extrapolate these contours to an integrated luminosity of 3000~$\text{fb}^{-1}$. To achieve this, we scale the ATLAS uncertainties for the current luminosity ($\text{Lumi}_{\text{ATLAS}} $) of 139 $\text{fb}^{-1}$ by a naive scaling factor
\begin{equation}
	\sigma_{\text{HL}}	= \frac{1}{\sqrt{\text{Lumi}_{\text{HL}}/\text{Lumi}_{\text{ATLAS}}}}\sigma_{\text{ATLAS}}\,.
	\footnote{We have ignored the individual scaling factors of the statistical and systematic uncertainties.}
\end{equation}
The projected HL-LHC uncertainties are tabulated in column 3 of Tab.~\ref{tab:exp_kappa} for an unchanged correlation matrix (column 4). We stress that we use this extrapolation purely to obtain a qualitative outlook for the HL-LHC, but note that this rescaling provides a good approximation of the comprehensive investigation of~\cite{deBlas:2019rxi}.
The resulting constraints at  68$\%$ and 95$\%$ confidence, assuming consistency with the SM, are shown in Fig.~\ref{fig:deltachi2_kappa_proj}.

As is visible from Fig.~\ref{fig:deltas}, we have rescaled the $\zeta_2$ axes to account for the loop suppression. Given that the correlation $\kappa_V(\kappa_{2V})$ is related to a weak radiative correction, the bounds on $\kappa_{2V}=1+\zeta_2$ from single Higgs data are loose, even when we consider the direct sensitivity of the early stage LHC programme shown in Fig.~\ref{fig:unitarity}. Extrapolations to $3/$ab show that single Higgs physics will greatly collapse the current limits along the $\zeta_1$ direction, however the loop suppression implies a sensitivity $|\zeta_2|\sim 4$.  

A coupling modifier of this size is already in tension with the current measurement results detailed in Sec.~\ref{sec:kap2v}. This shows that the $\kappa_{2V}$ and $\kappa_V$ are largely independent parameters at the LHC as is naively suggested by Eq.~\eqref{eq:heftlo}. 
This highlights the need to increase the sensitivity coverage to $\zeta_2$ through direct searches further, which we will turn to in Sec.~\ref{sec:GNN}.

%%%%%%%%%%%%%%%%%%%
\section{Enhancing direct collider sensitivity with Graph Neural Networks}
\label{sec:GNN}
%%%%%%%%%%%%%%%%%%%
\begin{figure}[!b]
\centering \includegraphics[width=11cm]{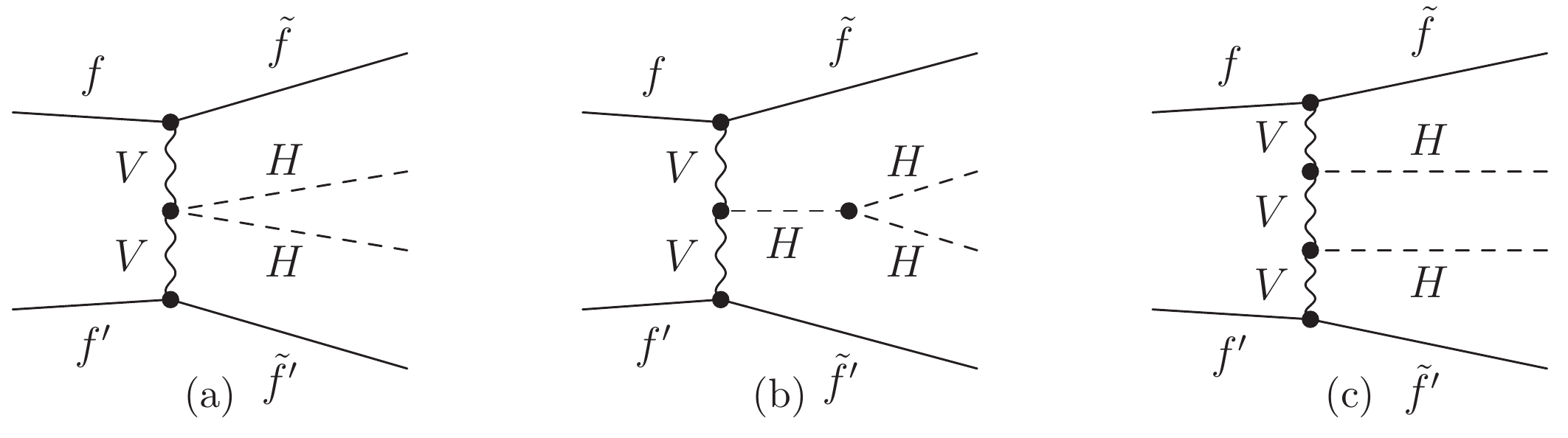}
\caption{\label{fig:wbf}  Representative Feynman diagrams contributing to WBF $p(f) p(f')\to HH j j$. We consider $H\to b\bar b$ final states, but do not include the decays to the diagrams.}
\end{figure}
%%%%%%%%%%%%%%%%%%%%%%
%
Higgs pair production via WBF has been discussed in Refs.~\cite{Dolan:2013rja,Dolan:2015zja,Bishara:2016kjn,Kilian:2018bhs,Arganda:2018ftn} (see also \cite{Cappati:2022skp} for a recent dimension 8 SMEFT study). The process shares the properties of single Higgs WBF production~\cite{Cahn:1983ip,Figy:2003nv,Jager:2006zc,Jager:2006cp,Ciccolini:2007ec} especially when we consider LHC-relevant QCD corrections~\cite{Figy:2008zd,Frederix:2014hta,Dreyer:2018qbw}. To select the events for the WBF topology,~Fig.~\ref{fig:wbf}, we choose two forwarded jets in opposite detector hemispheres (opposite signs of pseudorapidity $\eta_{j1}\cdot \eta_{j2} < 1$) with the invariant mass $m_{jj}$ larger than 500 GeV.  For the minimum transverse momentum $p_T$ of WBF jets, we choose 50 GeV. We select events with four $b$-jets in the central region $|\eta_{b-\text{jets}}| < $  2.5 with minimum $p_T\geq 20$ GeV.\footnote{We use a flat tagging efficiency of $70\%$ and ignore mistagging. For the EdgeConv GNN strategy we employ below, mistagging will predominantly change the background normalisation, which can be compensated through a tuned working point of the classifier. We comment on uncertainties from the background normalisation further below.}  Events are generated in proton-proton collisions at $\sqrt{s}=13$~TeV using \mg{}~\cite{Alwall:2014hca} at leading order precision {\reply{using the standard setting scale choices.}}\footnote{\reply{It is known that the NLO QCD corrections are relatively mild~(e.g. \cite{Figy:2003nv,Frederix:2014hta}) and can be largely captured through adapted weak-boson fusion factorisation scale choices inspired by the ``double-DIS'' structure of WBF~\cite{Jager:2006zc,Jager:2006cp}. We will see below that the used approach reproduces the current experimental outcome reasonably well to warrant a high-luminosity extrapolation.}} The main SM background for the $4b+2 j$  final state comes from multijet QCD processes~\cite{Dolan:2013rja,Dolan:2015zja,Bishara:2016kjn,Kilian:2018bhs,Arganda:2018ftn}. \reply{After the given VBF selection cuts, the QCD multijet background cross-section is 4.41 $pb$, while the signal cross-section is 0.086 $fb$, which highlights the necessity to efficiently reduce the background.}

Going beyond `traditional' techniques, utilising Graph Neural Networks (GNNs) is a motivated avenue. These strategies are tailored to accessing the graph structure of differential cross section data under realistic conditions that we typically parametrise through Feynman diagram calculations in theoretical reference calculations. Hence,
recent applications to particle physics data in a range of areas stretching from QCD~\cite{Dreyer:2020brq}, over anomaly detection~\cite{Atkinson:2021nlt} (including IR safe formulations~\cite{Konar:2021zdg,Atkinson:2022uzb}), to SMEFT parameter fits~\cite{Atkinson:2021jnj} have highlighted GNNs as versatile and highly sensitive tools for BSM discrimination. Given the special phenomenological properties of WBF processes, we can therefore expect similar performance improvements in this channel.

%%%%%%%%%%%%%%%%%%%%%%
\begin{figure}[!t]
\centering 
\subfloat[]{\includegraphics[width=6cm]{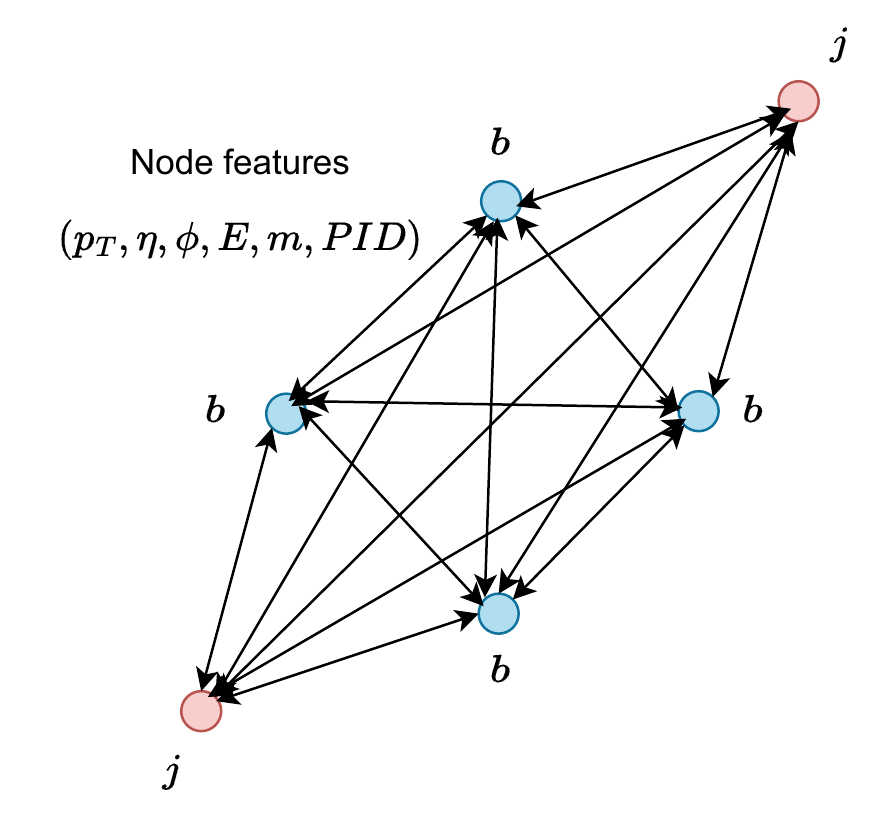}\label{fig:connect}}
\subfloat[]{\includegraphics[width=8cm]{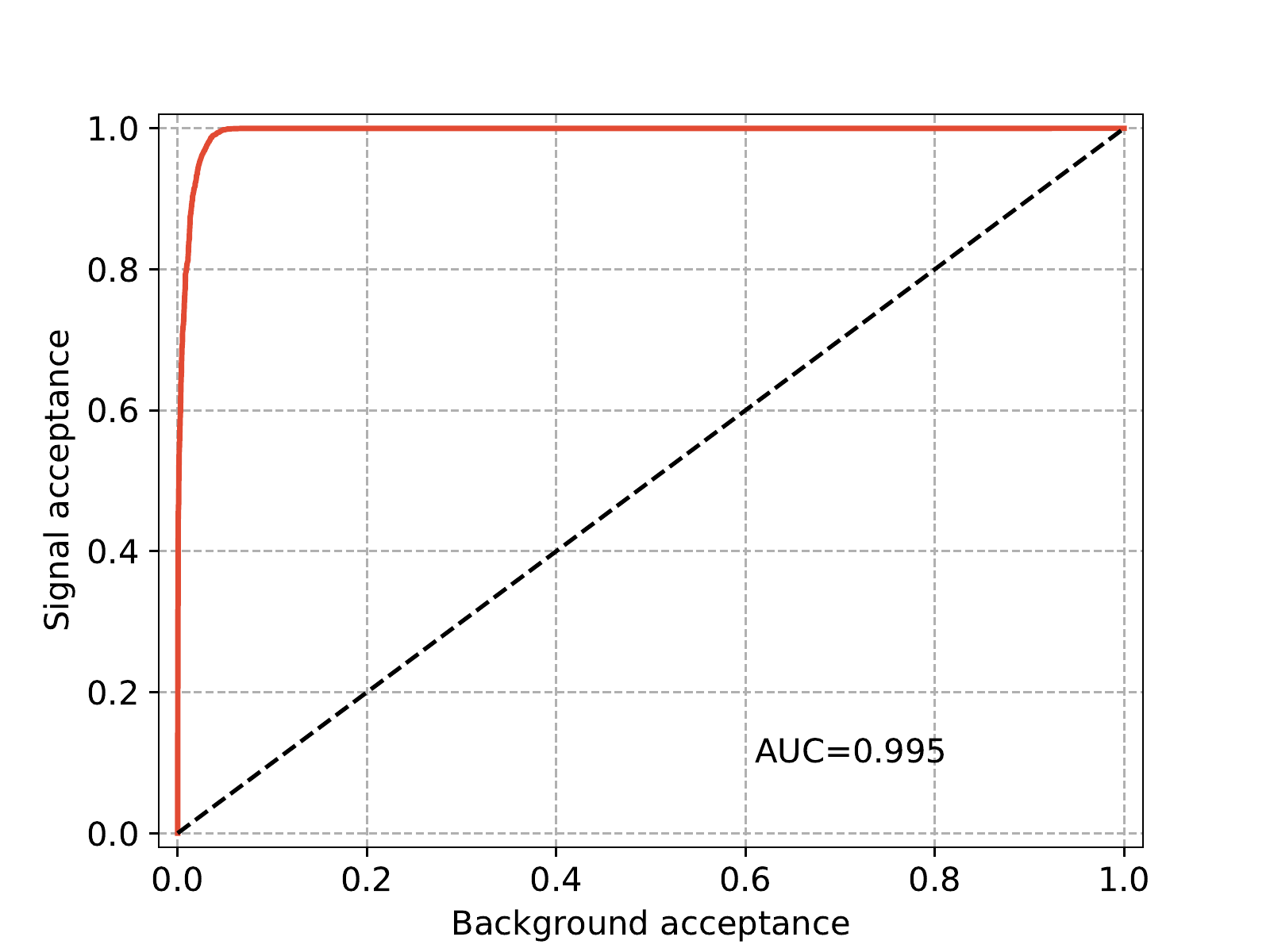}\label{fig:roc}}
\caption{\label{fig:Event}  A representative diagram of an event as a fully connected graph is shown on the left. The network performance is shown on the right with a receiver operating characteristic (ROC) curve along with the corresponding AUC=0.995.}
\end{figure}
%%%%%%%%%%%%%%%%%%%%%%

To construct a GNN for the present signal vs. background discrimination task, we use a fully-connected bi-directional graph to represent the $4b+ 2j$ event (Fig.~\ref{fig:connect}). Each jet is represented by a node, which is associated with a node feature vector $[p_T, \eta, \phi, E, m, PID]$ (representing transverse momentum, pseudorapidity, azimuthal angle, energy, mass and particle identification number, respectively). To implement the graph structure, we use the {\sc Deep Graph Library}~\cite{wang2020deep} and {\sc PyTorch}~\cite{paszke2019pytorch} and choose an Edge Convolution (EdgeConv) network to classify signal and background. Edge convolution is known to be particularly suited for extracting high-level variable information as edge features from given low-level node features~\cite{wang2019dynamic}. The message passing function for the Edge convolution is defined as 
%%%%%%%%%%%%%%%%%%%%%
\begin{equation}
	\label{eq:edge_conv}
	\vec{x}_{i}^{\,(l+1)} = \frac{1}{|\mathcal{N}(i)|} \sum_{j \in \mathcal{N}(i)} {\text{\sc{ReLU}}}\left(\Theta \cdot (\vec{x}_j^{\,(l)} - \vec{x}_i^{\,(l)}) + \Phi \cdot (\vec{x}_i^{\,(l)})\right)\,.
\end{equation}
%%%%%%%%%%%%%%%%%%%%%
Here, $\vec{x}_i^{\,(l)}$ represents the input node features of node $i$ in the $l$-th message passing layer, with $l=0$ denoting the input node features of the graph. The neighbourhood set $\mathcal{N}(i)$ consists of all nodes in the graph connected to node $i$. The linear layers $\Theta$ and $\Phi$ take the vectors $\vec{x}_j - \vec{x}_i$ and $\vec{x}_i$, respectively, and map them to the same vector space. We apply a single message passing step producing 40-dimensional node features.\footnote{Further details on GNNs, EdgeConv and architecture design can be found in Ref.~\cite{Atkinson:2021jnj}.} Since we aim to classify graphs, we apply a mean graph readout operation to these node features. A classifier multilayer perceptron with single layers having 40 nodes, with ReLU activation, takes these features and outputs a vector of two dimensions (after being properly normalised with a SoftMax activation function). We minimise the cross-entropy loss function using the {\sc{Adam}} optimiser~\cite{DBLP:journals/corr/KingmaB14} with a learning rate of $10^{-3}$. Given the distinct phenomenological properties, we also find that already a shallow network is well-suited to obtain good discrimination.

The network is trained for 100 epochs with batch sizes of 124 for a $\kappa_{2V}$-enriched parameter point sample ($\kappa_{2V},\kappa_V=2,1$). We split the total data into $80\%$, $10\%$ and $10\%$ for training, validation and testing, respectively. We verify that the validation and the training accuracies are comparable to avoid overtraining. The receiver operating characteristic curve of the network is shown in Fig.~\ref{fig:roc} with an area under curve (AUC) of 0.995, implying a very high classifier performance.\footnote{\reply{The AUC is slightly reduced to 0.941 when considering parton-showering and underlying event; this reduction can be compensated by a choice of working point, highlighting the phenomenological distinction of the signal and QCD background. Electroweak backgrounds that are not considered here, might require additional improvements.}}

%%%%%%%%%%%%%%%%%%%
\begin{figure}[!t]
	\centering 
	\subfloat[]{\includegraphics[width=7.4cm]{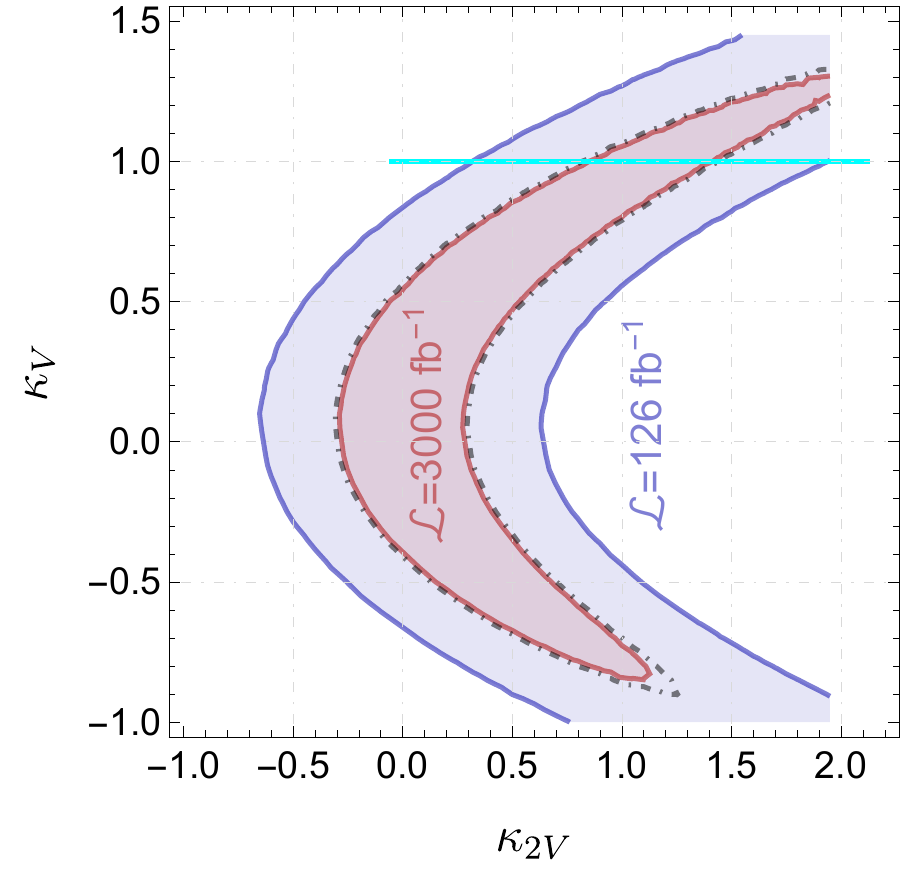}\label{fig:GNN_limit_kappa} }
	\hfill\subfloat[]
	{\includegraphics[width=7.4cm]{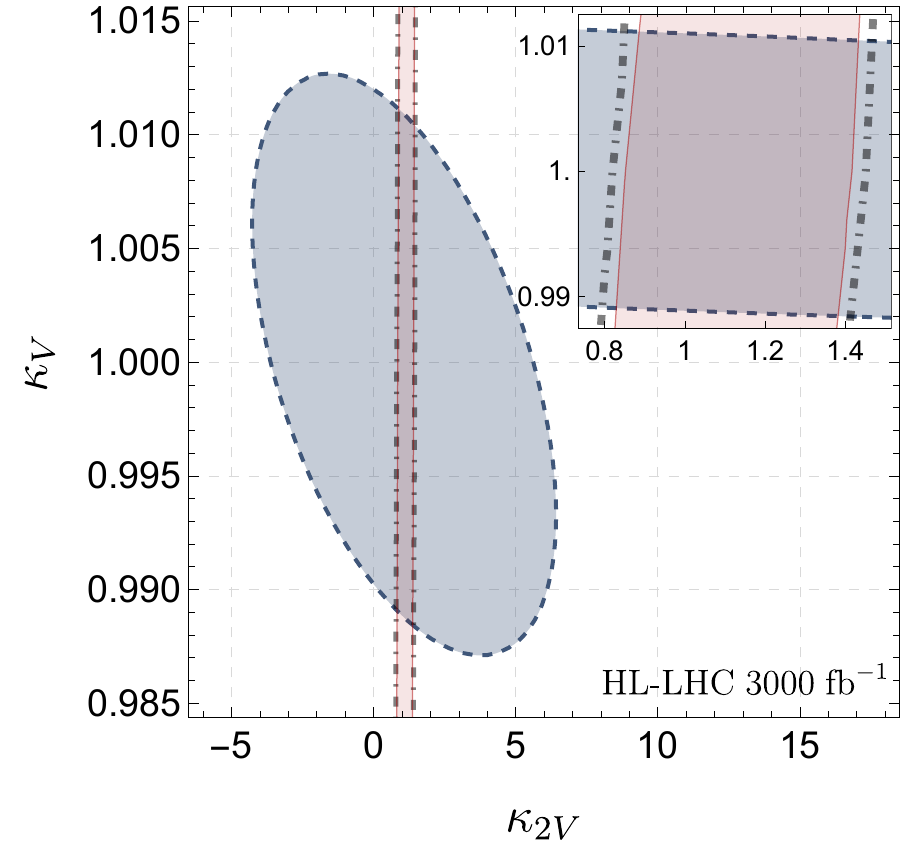}\label{fig:deltachi2_k2VkV_proj}} 
	\caption{ Exclusion limits on $\kappa_{2V}$ and $\kappa_V$ parameter space obtained from the GNN output trained for the $\kappa_{2V}$ enriched sample ($\kappa_{2V},\kappa_V = 2, 1$) are shown in Fig.~\ref{fig:GNN_limit_kappa}. The cyan line represents the limits obtained from the ATLAS analysis for the fixed value of $\kappa_V=1$~\cite{ATLAS:2022ycx}. The dot-dashed contour shows the effect of increasing the background by $25\%$. In Fig.~\ref{fig:deltachi2_k2VkV_proj}, HL-LHC projected constraints  are overlaid on the 95$\%$ C.L. region shown in Fig.~\ref{fig:deltachi2_kappa_proj} for $\kappa_{2V}$ and $\kappa_V$ parameter space. }
\end{figure}
%%%%%%%%%%%%%%%%%%%

With the classifier optimised as described above, we can obtain the limit on $\kappa_{2V}$ by fixing $\kappa_V = 1$ (we do not include modified trilinear Higgs interactions as in the previous sections) to compare the performance of the GNN with ATLAS analysis of Ref.~\cite{ATLAS:2022ycx}. \reply{For the optimal working point on the ROC curve, the QCD multijet background cross-section drops to is 0.88~fb, while the signal cross-section remains 0.037~fb. The exclusion limits are calculated using the significance $\sigma={N_s}/{\sqrt{N_s+N_b}}$, where $N_s$ is the number of signal events and $N_b$ is the number of background events at a particular luminosity.} This is shown in Fig.~\ref{fig:GNN_limit_kappa}, where we show the ATLAS constraint as a horizontal (cyan) line. Firstly, this agreement acts as a good validation of our procedure in comparison to the realistic experimental measurement, which gives us confidence for the HL extrapolation that we will perform next. The relative improvement suggested in Fig.~\ref{fig:GNN_limit_kappa} might be optimistic. Given the necessity to overcome large backgrounds, the choice of working points and the accuracy of the background simulation are factors that can impact the discrimination, in particular when we deal with the large statistics of the HL phase. The competitiveness of our idealised GNN analysis, however, demonstrates that such techniques deserve consideration as part of realistic experimental analyses. We will also comment on a range of potential architecture improvements that we have not included in our analysis below.

Using the $\kappa_{2V}=2$ as a reference point also for a more general scan, we choose a fixed working point of the classifier which maximises the significance to obtain exclusion contours in the $\kappa_{2V},\kappa_V$ parameter space also shown in Fig.~\ref{fig:GNN_limit_kappa} (this includes the changes to the $H\to b \bar b$ branching ratio as a function of $\kappa_V$).
To compare the level of sensitivity, we overlay these, projected to 3/ab HL-LHC target with the expected sensitivity from single Higgs measurements detailed in Sec.~\ref{sec:constraints} in Fig.~\ref{fig:deltachi2_k2VkV_proj}. Note that in this exploratory comparison, we have not included subdominant background from weak multi-boson processes or top production, which will additionally degrade the sensitivity~(see e.g.~\cite{Dolan:2015zja}). These processes are, however, also characterised by different kinematic and QCD radiation properties, and we can expect significant discrimination when considering these additional contributions in the GNN classification.  To get an estimate of how additional backgrounds modify the direct search sensitivity, we modify the considered dominant QCD through a naive shifting by 25\% in normalisation. Including these to Fig.~\ref{fig:GNN_limit_kappa}, we see that there is minor change in the sensitivity, and good sensitivity to $\kappa_{2V}$ can be achieved. 

The GNN could be further improved through a more dedicated inclusion of $\kappa_{2V}-\kappa_{V}$ correlations to the classification, e.g., through a multi-class network architecture discussed in Ref.~\cite{Atkinson:2021jnj}, which could also be extended to subdominant background contributions to specifically combat those when the QCD contribution has been removed sufficiently. A more robust graph embedding could help in such an approach as well. While the QCD background phenomenology is very different from the WBF signal (highlighted by the fact that a shallow network is sufficient for very good discrimination), different electroweak correlation structures such as top and weak boson decays impart a more tree-like structure that could be further exploited when these become relevant. Insensitivity to differential distributions of the node features, as well as overall normalisations, could be achievable using adversarial networks~\cite{Louppe:2016ylz}. We leave more detailed investigations for future work.

%%%%%%%%%%%%%%%%%%%
\section{Summary and Conclusions}
\label{sec:conc}
%%%%%%%%%%%%%%%%%%%
In this work, we have presented a comprehensive and theoretically consistent discussion of $\kappa_V-\kappa_{2V}$ correlations by employing Higgs Effective Field Theory, which puts related analyses at the LHC~\cite{ATLAS:2020jgy,ATLAS:2022ycx,CMS:2022hgz} on a theoretically firm footing. We show that single Higgs constraints can be formulated for $\kappa_{2V}$ analyses, however, given that these effects arise as a weak radiative correction, the single-Higgs $\kappa_V$ constraints are relatively loose and not competitive when compared to the LHC's sensitivity to $\kappa_{2V}$ already at this stage of the programme. Indeed, the LHC's sensitivity pattern when mapped onto the HEFT Lagrangian~Eq.~\eqref{eq:heftlo} and its weak corrections allows us to treat different Higgs interactions in the gauge sector as largely independent parameters.

As $\kappa_{2V}$ analyses are mainly driven by the direct investigation of weak $HHjj$ production, an enhanced direct sensitivity of future LHC runs provides the best motivated avenue to obtain a more fine-grained picture of the Higgs boson's gauge interactions along these lines. To this end, we employ Graph Neural Network techniques to demonstrate that the direct sensitivity to $\kappa_{2V}$ can be enhanced as GNNs formidably exploit the particular structure of the decay, colour and kinematical correlations of the $HHjj$ signal compared to the dominant QCD backgrounds. Improvements to $\kappa_{2V}\sim 1\pm 40\%$ could then be within the reach of the HL-LHC, in particular as improvements through multi-class GNNs and additional architecture improvements are likely to add further sensitivity.

%%%%%%%%%%%%%%%%%%%
\section*{Acknowledgements}
%%%%%%%%%%%%%%%%%%%
CE thanks Bill Murray for helpful discussions. 
The work of A. is funded by a Leverhulme Trust Research Project Grant RPG-2021-031. O.A. is supported by the UK Science and Technology Facilities Council (STFC) under grant ST/V506692/1. A.B. is funded by the STFC under grant ST/T000945/1. C.E. acknowledges funding by the Leverhulme Trust (RPG-2021-031), by the STFC under grant ST/T000945/1, and by the Institute of Particle Physics Phenomenology Associateship Scheme. P.S. is funded by an STFC studentship under grant ST/T506102/1.

%%%%%%%%%%%%%%%%%%%%%%%%%
\appendix
\section{UV divergent parts of renormalisation constants}
\label{sec:uvdiv}
%%%%%%%%%%%%%%%%%%%%%%%%%
In this section, we provide the explicit form of the UV divergent parts renormalisation constants used to renormalise the Lagrangian of Sec.~\ref{sec:heftlag} required in this work. The field strength renormalisations are obtained by evaluating all relevant (off-shell, one-particle irreducible) two-point function $X\to Y$ contributions derived from Eq.~\eqref{eq:heftlo}, which we represent as
\begin{equation}
\parbox{2.5cm}{\vspace{0.2cm}\includegraphics[width=2.5cm]{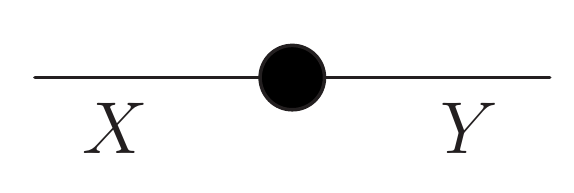}}\,,
\end{equation}
to be contrasted with 2-point insertions from Eq.~\eqref{eq:chir4op} represented by
\begin{equation}
\parbox{2.5cm}{\vspace{0.2cm}\includegraphics[width=2.5cm]{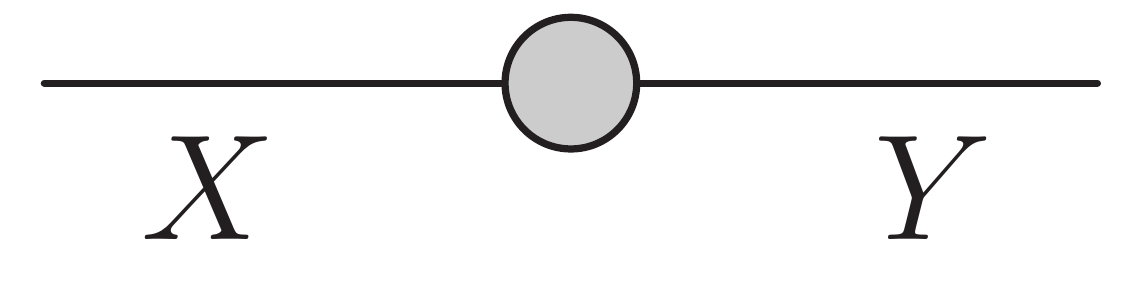}}\,.
\end{equation}
The counter terms are then pictorially determined from the renormalisation conditions (RCs) through
\begin{equation}
\left[\left(\parbox{2.5cm}{\vspace{0.2cm}\includegraphics[width=2.5cm]{sminser.pdf}}+
\parbox{2.5cm}{\vspace{0.3cm}\includegraphics[width=2.5cm]{d6inser.pdf}}\right)
+\parbox{2.5cm}{\vspace{0.2cm}\includegraphics[width=2.5cm]{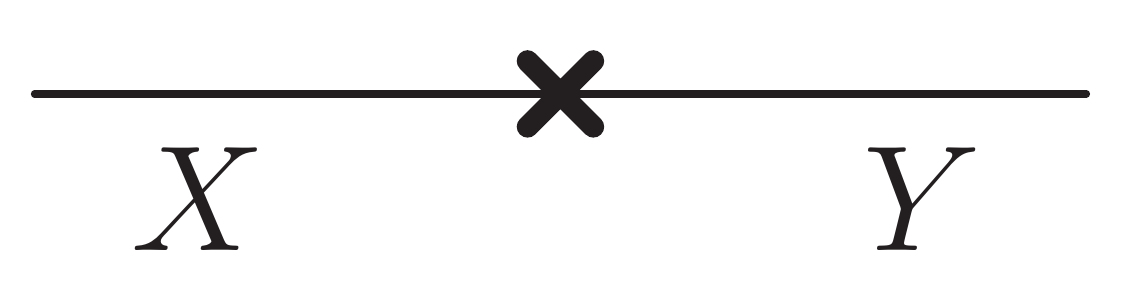}}\right]_{\text{RCs}}=0\,.
\end{equation}
The counter term diagrams will typically contain terms from Eq.~\eqref{eq:chir4op} which we highlight via 
\begin{equation}
\delta \left[\parbox{2.5cm}{\vspace{0.2cm}\includegraphics[width=2.5cm]{d6inser.pdf}}\right] \,.
\end{equation}

In the conventions detailed above, the contribution to the off-shell Higgs boson two-point function $H(q)\to H(q)$ that  needs to be included is
\begin{equation}
\parbox{2.5cm}{\vspace{0.2cm}\includegraphics[width=2.5cm]{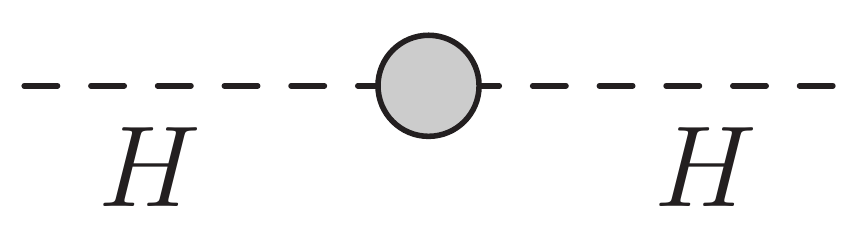}} = {e^2\over 4 m_W^2 s_W^2} q^4 a_{\Box\Box}  \,.
\end{equation}
The counter term, including the ${\cal{L}}_{4}$ contributions, can then be written as
\begin{equation}
\parbox{2.5cm}{\vspace{0.2cm}\includegraphics[width=2.5cm]{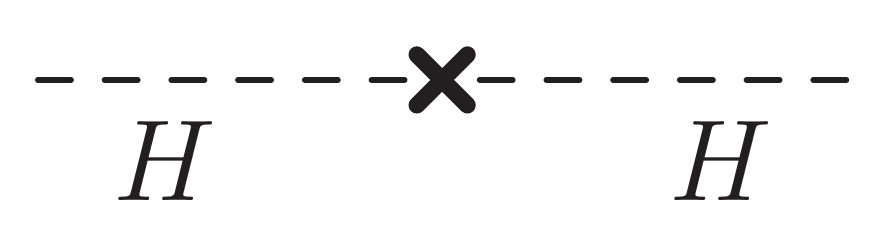}} = \delta Z_H (q^2 - M_H^2) +\delta M_H^2 + \delta \left[ 
\parbox{2.5cm}{\vspace{0.2cm}\includegraphics[width=2.5cm]{hd6.pdf}} \right]\,,
\end{equation}
with divergent parts given by
\begin{equation}
\begin{split}
\delta Z_H\big|_\Delta =& {3\alpha  \over 16\pi c_W^2 s_W^2 } (3-2s_W^2)(1+\zeta_1)^2  -  {3\alpha  \over 8\pi  s_W^2} {m_t^2+m_b^2\over M_W^2} \,,\\
\delta M_H^2\big|_\Delta =& {3\alpha\over 32\pi M_W^2 s_W^2} (3\kappa_3^2+\kappa_4) M_H^4 + {3\alpha\over 16\pi c_W^2 s_W^2} M_H^2 (2 s_W^2-3)(1+\zeta_1)^2 
\\ & +{3\alpha\over 16\pi c_W^4s_W^2} M_W^2 (3-4s_W^2+2s_W^4) ( 3+2\zeta_1(2+\zeta_1)+\zeta_2) \\ 
&+ {3\alpha \over 8\pi M_W^2 s_W^2 } M_H^2 \left( m_b^2+m_t^2 \right)  - {9\alpha \over 4\pi M_W^2s_W^2} (m_t^4+m_b^4)\,,\\
\delta a_{\Box\Box}\big|_\Delta =&  -{3\over 64\pi^2} (1+\zeta_1)^2\,.
\end{split}
\end{equation}
Note that the Higgs renormalisation is manifestly gauge-independent as the Higgs is a singlet field in HEFT~\cite{Herrero:2021iqt} (this also applies to the tadpole).
The transverse part of the $W(q) \to W(q)$ boson polarisation function receives no contributions from Eq.~\eqref{eq:chir4op}
\begin{equation}
\left[\parbox{2.5cm}{\vspace{0.2cm}\includegraphics[width=2.5cm]{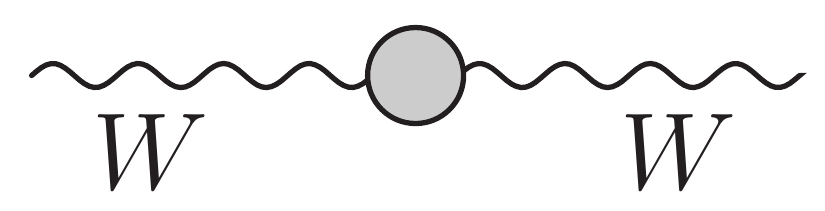}}\right]_T = 0 \,,
\end{equation}
leading to a counter term insertion
\begin{equation}
\left[\parbox{2.5cm}{\vspace{0.2cm}\includegraphics[width=2.5cm]{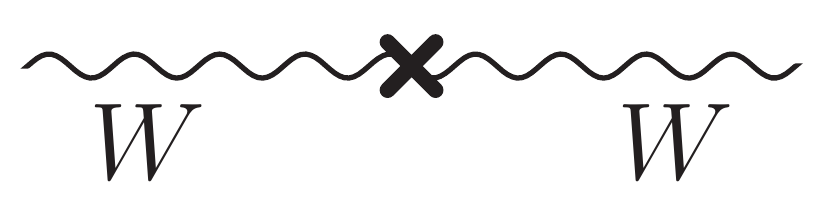}} \right]_T= -\dzw (q^2 - M_W^2)  + \dmw \,,
\end{equation}
with
\begin{equation}
\begin{split}
\dzw \big|_\Delta =& - {\alpha \over 48\pi s_W^2} \left( 6\xi_W + 6s_W^2(\xi_A-\xi_Z)+6\xi_Z+\zeta_1(2+\zeta_1) -2 \right) \,,\\
\dmw\big|_\Delta =&  -\frac{{\alpha}}{48 \pi  c_W^2 s_W^2} \big(3 c_W^2 (6 (m_b^2+m_t^2)+M_H^2 ({\zeta_1}
   ({\zeta_1}+2)- {\zeta_2})) \\ 
   &-10 {M_W^2}
   c_W^2 {\zeta_1} ({\zeta_1}+2)+{M_W^2} (11-20
   s_W^2)\big)\,.
\end{split}
\end{equation}
Similarly, we find for transverse $Z$ boson polarisation from $Z(q)\to Z(q)$
\begin{equation}
\left[\parbox{2.5cm}{\vspace{0.2cm}\includegraphics[width=2.5cm]{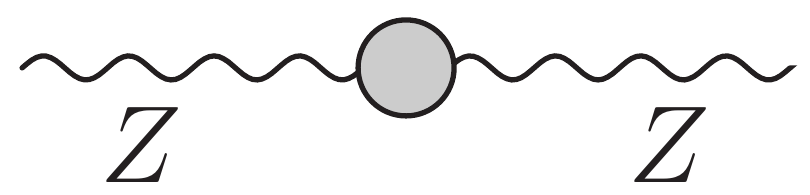}}\right]_T = -2 {e^2\over c_W^2} M_Z^2 a_0 - 2 {e^2} q^2 a_1 \,,
\end{equation}
and
\begin{equation}
\left[\parbox{2.5cm}{\vspace{0.2cm}\includegraphics[width=2.5cm]{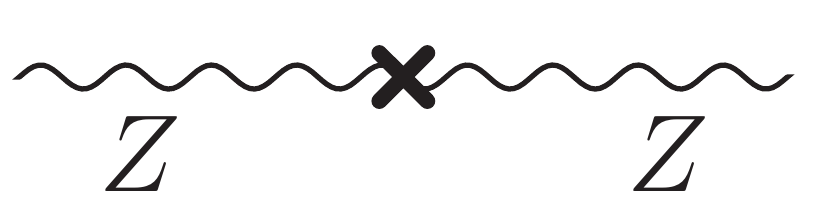}} \right]_T= -\dzzz (q^2 - M_Z^2)  +\dmz  + \delta \left[ 
\parbox{2.5cm}{\vspace{0.2cm}\includegraphics[width=2.5cm]{zd6.pdf}} \right]_T
  \,,
\end{equation}
the expressions
\begin{equation}
\begin{split}
 \dzzz\big|_\Delta =&- {\alpha\over 48\pi s_W^2 c_W^2} \big(12 \xi_W -2 + 4 s_W^2(1-6\xi_W+s_W^2(20+3\xi_W)) \\& + \zeta_1(2+\zeta_1)\big)  -2 e^2 \delta a_1 \,,\\
\dmz\big|_\Delta =&  \frac{ \alpha } {48 \pi 
   {c_W^4 s_W^2}} \left(-3 c_W^2 ({M_H^2}
   ({\zeta_1} (2+{\zeta_1})-{\zeta_2}) + 6 ({m_b^2+m_t^2}) )\right. \\ & + \left. {M_W^2} \left(44
   s_W^4+58 s_W^2+10 {\zeta_1} (2+{\zeta_1})-11\right)\right) + 2 {e^2\over c_W^2} M_Z^2 (\delta a_0 + c_W^2 \delta a_1)\,, \\
\delta a_0\big|_\Delta =& -{9\over 128 \pi^2} (\zeta_1^2-2\zeta_1)\,,\\
\delta a_1\big|_\Delta =& -{1\over 192 \pi^2} (\zeta_1^2-2\zeta_1) \,.
\end{split}
\end{equation}
For the $\gamma-\gamma$ polarisation, we obtain
\begin{equation}
\left[\parbox{2.5cm}{\vspace{0.2cm}\includegraphics[width=2.5cm]{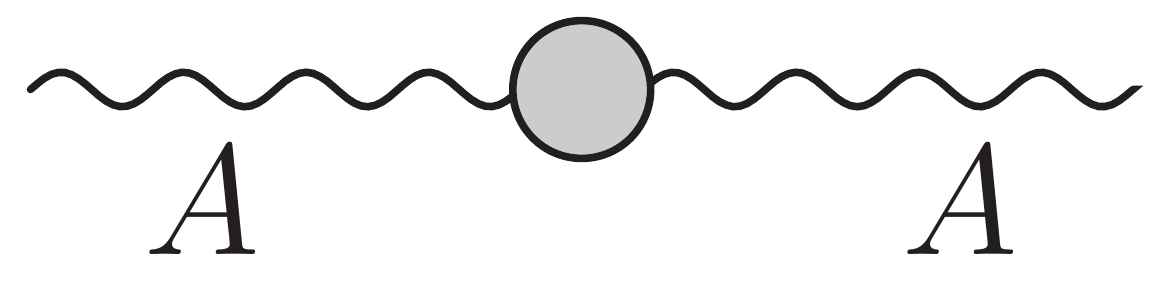}}\right]_T = 
2e^2 q^2 a_1\,,
\end{equation}
providing
\begin{equation}
\left[\parbox{2.5cm}{\vspace{0.2cm}\includegraphics[width=2.5cm]{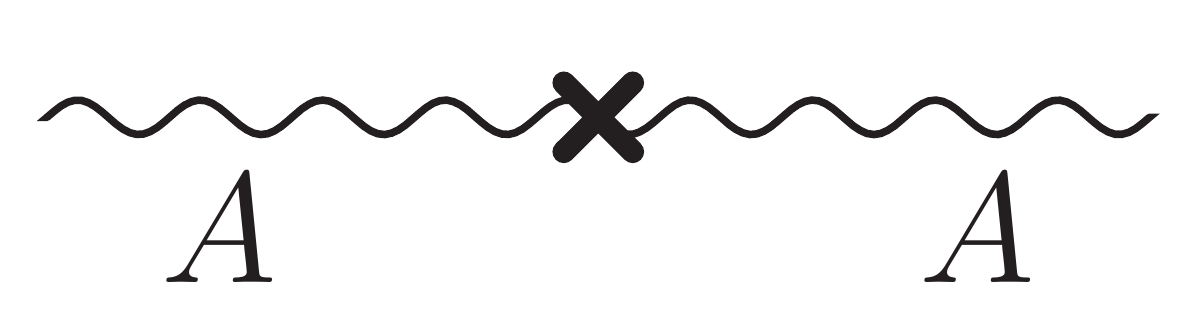}} \right]_T= -\dzaa q^2   + \delta \left[ 
\parbox{2.5cm}{\vspace{0.2cm}\includegraphics[width=2.5cm]{ad6.pdf}} \right]_T\,.
\end{equation}
leading to
\begin{equation}
\begin{split}
 \dzaa\big|_\Delta =& -{\alpha \over 12\pi}(20+3\xi_W) \,.\\
\end{split}
\end{equation}
Finally, the $Z(q)\to\gamma(q)$ mixing is
\begin{equation}
\left[\parbox{2.5cm}{\vspace{0.2cm}\includegraphics[width=2.5cm]{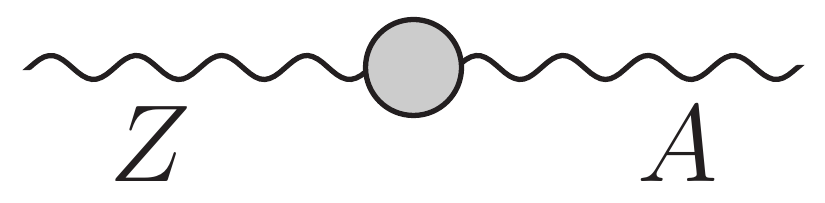}}\right]_T =  - {e^2\over s_W c_W}(c_W^2-s_W^2) a_1\,,
\end{equation}
leading to 
\begin{equation}
\left[\parbox{2.5cm}{\vspace{0.2cm}\includegraphics[width=2.5cm]{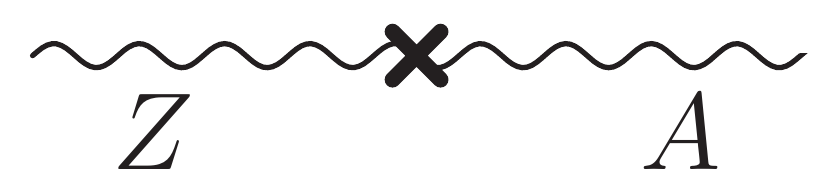}}\right]_T = -(\dzaz+\dzza) q^2 + {\dzza\over 2} M_Z^2 
+ \delta\left[\parbox{2.5cm}{\vspace{0.2cm}\includegraphics[width=2.5cm]{zad6.pdf}}\right]_T 
\end{equation}
with
\begin{equation}
\begin{split}
 \dzaz\big|_\Delta =& -{\alpha \over 12 \pi c_W s_W } \left ( 10-3\xi_W + s_W^2(31+3\xi_W) \right) - 2 {e^2\over c_W s_W} (s_W^2-c_W^2) \delta a_1 \,,\\
\dzza\big|_\Delta =&  {\alpha \over 4\pi } {c_W\over s_W} (3+\xi_W) \,.
\end{split}
\end{equation}
These results agree with Ref.~\cite{Herrero:2021iqt} when mapped onto the conventions used there. In particular, the bare quantities agree, as do the relevant unrenormalised three-point functions of Sec.~\ref{sec:ren}. It is worth highlighting that the mass renormalisation constants are gauge-independent

%%%%%%%%%%%%%%%%%%%
%\bibliographystyle{JHEP}
\bibliography{paper}
%%%%%%%%%%%%%%%%%%%

\end{document}